\documentclass[a4paper,11pt]{article}
\usepackage{jinstpub}
\usepackage{subcaption}
\usepackage{booktabs}
\usepackage{lineno}
\usepackage{graphicx}   % 图片
\usepackage{subcaption} % 子图

\title{A High-Precision Clock Synchronization System for the CEPC Accelerator}
\author[a,b,c,1]{Jun Hu\note{Corresponding author.}}
\author[a,b]{Xin Zhou}
\author[a,b,c]{Xiaoshan Jiang}
\author[a,b,c]{and Dapeng Jin}

\affiliation[a]{Institute of High Energy Physics, Chinese Academy of Sciences,\\19B Yuquan Road, Shijingshan District, Beijing 100049, China}
\affiliation[b]{University of Chinese Academy of Sciences,\\19A Yuquan Road, Shijingshan District, Beijing 100049, China}
\affiliation[c]{Center for High Energy Physics, Henan Academy of Sciences,\\No. 228 Chongshi Lane, Zhengdong New District, Zhengzhou 450002, Henan, China}

\emailAdd{hujun@ihep.ac.cn}

\abstract{
The Circular Electron Positron Collider (CEPC) distributes a reference clock distributed to 192 control nodes along its 100~km underground tunnel. The required synchronization precision is 30~ps (standard deviation). We present an enhanced White Rabbit (WR)-based clock synchronization system designed to meet this requirement. A noise-budget analysis of the standard WR slave loop identifies the analog actuation chain (DAC + VCXO + multiplier PLL) and restart-induced timing uncertainty as the dominant limitations. In our redesigned node, the DAC+VCXO chain is replaced by a Si5345A DSPLL clock generator with DCO-based phase control, removing the board-level analog tuning stage. GTX transceiver phase alignment and manual byte-alignment fixing reduce restart uncertainty from 88.8~ps to 12~ps peak-to-peak. For multi-node operation, we introduce a cascaded global-control architecture with PC-side PID auto-tuned by TD3 reinforcement learning, on-chip-temperature feed-forward calibrated to $-0.76\,\mathrm{ps}/^\circ\mathrm{C}$. The measured point-to-point synchronization precision is 3.38~ps over 1~m fiber and 3.92~ps over 50~km. In a 12-level cascade, the end-node precision reaches 6.66~ps at constant temperature and 7.30~ps under a 13$\,^\circ$C temperature swing. Synchronized-clock TIE jitter stays below 1~ps regardless of cascade depth. Restart uncertainty is 2.82~ps (std.\ dev.). A 4-level cascade operated stably for 25 hours of continuous monitoring. All measured metrics fall well within the CEPC 30~ps budget.
}


\begin{document}
\maketitle
\flushbottom

\section{Introduction}\label{sec:intro}
% Section 1: Introduction

The Circular Electron Positron Collider (CEPC) is a proposed next-generation Higgs factory in China, designed for high-precision measurements of the Higgs boson and Standard Model particles at centre-of-mass energies of 91--360~GeV~\cite{ref_new27,ref_new28}. The accelerator complex consists of a linac, a damping ring, a booster and a 100~km collider ring housed in an underground tunnel at approximately 100~m depth. Along the tunnel, 192 control nodes are uniformly distributed in 192 side caverns; eight vertical shafts provide cabling and infrastructure access. Beam diagnostics and the RF system require their timing references to be aligned within a standard deviation of 30~ps between any two nodes, so this 30~ps budget applies to the timing system as well. The synchronization signal under study is a 62.5~MHz reference clock.

Large accelerators have historically adopted two distinct timing-distribution paradigms. \emph{Event-distribution} systems, such as the Micro-Research Finland (MRF) family deployed at LCLS, SwissFEL and HEPS~\cite{ref_new29,ref_new31,ref_new35}, broadcast encoded event codes over a fan-out tree; they deliver sub-nanosecond trigger precision but depend on counter-based delay compensation. \emph{Time-distribution} systems instead establish a common time-of-day at every endpoint, with White Rabbit (WR)~\cite{ref_new37,ref_new38} now widely adopted by next-generation facilities. WR integrates Synchronous Ethernet (Sync-E) for frequency transfer, IEEE~1588 PTPv2 for time transfer, and Digital Dual Mixer Time Difference (DDMTD) for picosecond-resolution phase measurement. It has been deployed at CERN's accelerator timing upgrade~\cite{ref_new2001}, and at SHINE, CSNS, HIAF, LHAASO, JUNO and SKA~\cite{ref_new39,ref_new40,ref_new41,ref_new44,ref_new45,ref_new46}. The CERN/LHC deployment originally targeted sub-nanosecond accuracy and sub-100~ps precision over 10~km links with up to 4 cascade levels~\cite{ref_new42}; typical reported standard WR precision is a few tens of picoseconds~\cite{ref_new83,ref_new87}. Meeting 30~ps precision across a 100~km ring with 192 nodes, however, demands improvements to the WR baseline on several fronts.

%\begin{figure}[!htbp]
%    \centering
%    \includegraphics[width=0.6\textwidth]{figures/image9.png}
%    \caption{Schematic illustration of the fiber-based clock synchronization principle from the master node to one downstream slave node. The %full CEPC deployment uses a multi-level cascaded topology built on this point-to-point link, as detailed in Section~\ref{sec:global}.}
%    \label{image9}
%\end{figure}

This paper presents a comprehensively enhanced WR-based clock synchronization system. Our main contributions are:

\begin{enumerate}
    \item A noise-budget analysis of the standard WR slave control loop that identifies the actuation chain (digital-to-analog converter (DAC) + voltage-controlled crystal oscillator (VCXO) + frequency-multiplying PLL), the reference-side clock-and-data-recovery (CDR) noise, and the DDMTD measurement chain as the dominant contributors, together with a separate characterization of restart-induced timing uncertainty.
    \item A redesigned clock synchronization node in which the external DAC+VCXO actuation chain is replaced by a Si5345A jitter-attenuating clock generator with an integrated digital signal-processing PLL (DSPLL) and digitally-controlled-oscillator (DCO) phase control, removing the board-level analog tuning node and achieving <1 ps TIE jitter with 3.38~ps point-to-point synchronization precision over 1~m fiber and 3.92~ps over 50~km.
    \item A firmware solution to the restart uncertainty problem based on gigabit transceiver (GTX) TX/RX phase alignment, manual byte-alignment fixing and a dedicated hardware initialization sequence, reducing restart uncertainty from a reference baseline of 88.8~ps to 12~ps peak-to-peak.
    \item A multi-level cascaded global control architecture with PC-side PID control whose gains are auto-tuned by twin-delayed deep deterministic policy gradient (TD3) reinforcement learning, combined with on-chip-temperature feed-forward compensation, delivering 6.66~ps precision over a 12-level cascade.
\end{enumerate}

The rest of this paper is organized as follows. Section~\ref{sec:wr} analyzes the noise budget of the standard WR slave loop and the restart-uncertainty limitation. Section~\ref{sec:hw} presents the hardware optimization of the clock node, replacing the DAC+VCXO actuation chain with the Si5345A DSPLL. Section~\ref{sec:fw} presents the firmware optimization, including the DCO control logic and the restart-uncertainty solution. Section~\ref{sec:global} describes the cascaded global control architecture, the temperature feed-forward calibration, and the RL-based PID auto-tuning method. Section~\ref{sec:results} reports the experimental results, including ablations against the standard WR baseline and a long-term run, and Section~\ref{sec:conclusion} concludes the paper.

\section{White Rabbit Technology and Performance Limitations}\label{sec:wr}
% Section 2: White Rabbit Technology and Performance Limitations

\begin{figure}[htbp]
    \centering
    \includegraphics[width=0.6\textwidth]{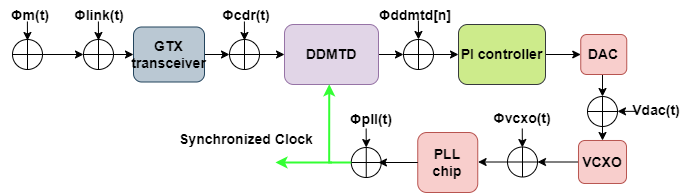}
    \caption{Main noise sources in the WR slave-node local synchronized-clock control loop. Reference-side and DDMTD measurement-chain noise are low-pass-shaped by $|H(f)|^2$; DAC, VCXO and multiplier-PLL noise enter through the actuation chain and are high-pass-shaped by $|H_e(f)|^2$.}
    \label{fig:noise_sources}
\end{figure}

To identify where optimization is most effective, we have systematically analyzed the WR slave-node control loop. In the standard WR architecture, a slave maintains synchronization through a closed-loop system (Main PLL, MPLL) comprising a DDMTD phase detector, PI controller, DAC, VCXO, and PLL frequency multiplier~\cite{ref_new37,ref_new38,ref_new2011}. The noise sources can be classified into three categories based on their injection points. \emph{Reference-side noise} ($S_{\phi,\mathrm{ref}}$) includes CDR-recovered clock phase noise (approximately $-97$~dBc/Hz at 1~Hz offset for Virtex-6 GTX~\cite{ref_new83}), master reference generation noise, and link noise; these are shaped by the closed-loop transfer function $|H(f)|^2$ which has low-pass characteristics. \emph{Measurement-chain noise} arises from the DDMTD phase detector, with a flicker floor around $-100$~dBc/Hz at 10~Hz and a white noise floor of approximately $-108$~dBc/Hz~\cite{ref_new83}, also low-pass-shaped. \emph{Actuation-chain noise} is dominated by the DAC (quantization plus thermal noise), the VCXO inherent phase noise ($-30$~dBc/Hz at 1~Hz), and the multiplier PLL (27.4~ps RMS jitter for CDCM61004). These enter through the error transfer function $|H_e(f)|^2$, whose high-pass character provides only $\sim-48$~dB suppression at 1~Hz, allowing significant low-frequency VCXO noise to leak through. The corresponding injection points within the WR slave-node control loop are illustrated in Fig.~\ref{fig:noise_sources}, and Table~\ref{tab:noise_budget} summarizes the three categories, the components involved, and their loop shaping. The actuation chain is identified as the dominant noise contributor, followed by the reference-side CDR and measurement-chain noise.

\begin{table}[htbp]
    \centering
    \caption{Noise budget of the standard WR slave-node control loop and the corresponding optimizations adopted in this work.}
    \label{tab:noise_budget}
    \begin{tabular}{lll}
        \toprule
        Category & Dominant components & Loop shaping \\
        \midrule
        Reference-side    & GTX CDR, master reference, link    & low-pass $|H(f)|^2$    \\
        Measurement chain & DDMTD phase detector               & low-pass $|H(f)|^2$    \\
        Actuation chain   & DAC, VCXO, multiplier PLL          & high-pass $|H_e(f)|^2$ \\
        \bottomrule
    \end{tabular}
\end{table}

A second critical limitation is the synchronization uncertainty upon device restart: the transmit and receive fixed delays ($\Delta_{\mathrm{TX}}$, $\Delta_{\mathrm{RX}}$), though assumed constant after calibration, can change after reset, power cycling, or link re-establishment. Measured values in a standard WR switch--node setup~\cite{ref_new87} show a restart-uncertainty standard deviation of 14.7~ps and peak-to-peak of 88.8~ps, making this a primary bottleneck for high-precision applications. Four independent GTX-level mechanisms are responsible: (i) TXOUTCLK phase ambiguity among the $N$ possible division states; (ii) TX phase-adjust FIFO read/write pointer variation; (iii) RX byte-alignment (Bit-Slide) variation; and (iv) RX elastic-buffer pointer variation. These mechanisms are analyzed in detail and addressed by targeted firmware countermeasures in Section~\ref{sec:fw}, which will be shown in Section~\ref{sec:results} to reduce the net effect by roughly a factor of 7.

The actuation-chain noise and the restart-uncertainty problem together set the agenda for the hardware and firmware work that follows. (Sections~\ref{sec:hw} and~\ref{sec:fw}).

\section{Hardware Optimization of the Clock Synchronization Node}\label{sec:hw}
% Section 3: Hardware Optimization of Clock Synchronization Node

Following the noise budget of Section~\ref{sec:wr}, this section addresses the dominant actuation-chain noise by replacing the external DAC+VCXO chain with an on-chip DSPLL-based clock generator. The surrounding power and clocking circuits are also reworked to further reduce the reference-side and measurement-chain noise contributions listed in Table~\ref{tab:noise_budget}.

\subsection{Si5345A Clock Generator and DSPLL Architecture}

The main hardware change is replacing the external DAC+VCXO actuation chain with a Silicon Labs Si5345A jitter-attenuating clock generator. The Si5345A integrates a fourth-generation DSPLL (Digital Signal Processing Phase-Locked Loop) architecture with MultiSynth\texttrademark{} fractional frequency synthesis technology~\cite{ref_new88}. The key characteristics relevant to our application are: differential input frequency range 8~kHz--750~MHz, output range 100~Hz--1028~MHz; typical output jitter of 90~fs RMS (12~kHz--20~MHz); programmable loop bandwidth from 0.1~Hz to 4~kHz; DCO mode with fine frequency/phase control (minimum step $\sim$0.5~ps); and 10 independently configurable output channels supporting LVDS, LVPECL, LVCMOS, CML, and HCSL signaling.

The DSPLL architecture transforms the traditional analog PLL control entirely into the digital domain. The phase detector digitizes the input signal at the front end, the loop filter is implemented digitally (eliminating external passive components), and the DCO provides high-resolution digital frequency control. Integrated on-chip LDOs and phase error cancellation circuitry provide superior noise immunity.

In our system, the Si5345A is configured with a custom 50~MHz low-noise crystal oscillator (phase noise: $-115$~dBc/Hz at 10~Hz, $-125$~dBc/Hz at 100~Hz, $-130$~dBc/Hz at 1~kHz, $-140$~dBc/Hz at 10~kHz and $-145$~dBc/Hz at 100~kHz; frequency stability $\pm0.5$--$1.0$~ppm/year and $\pm0.5$--$1.0$~ppm over $-40$ to $+85\,^\circ$C) connected to the XA/XB inputs. The DCO mode is enabled on the output dividers N0 and N1 with a configured frequency step of 0.5~ppm and a tuning range of $\pm200$~ppm. The intrinsic DCO phase resolution corresponds to $\sim$0.5~ps at the 62.5~MHz output; in our firmware implementation a longer dwell time of 2~$\mu$s per step is used to satisfy the Si5345A 1~$\mu$s minimum interval between consecutive FINC/FDEC toggles, yielding an effective control step of 1~ps (Section~\ref{sec:fw}). The output clock mapping is summarized in Table~\ref{tab:si5345a_out}.

\begin{table}[htbp]
    \centering
    \caption{Si5345A output clock configuration. The other five output channels are not used in this design.}
    \label{tab:si5345a_out}
    \begin{tabular}{cccc}
        \toprule
        Port & Standard & Frequency & Purpose \\
        \midrule
        OUT0 & LVDS 1.8V & 62.5~MHz & Local synchronized clock for FPGA logic \\
        OUT1 & LVDS 1.8V & 125~MHz & GTX transceiver \emph{TX} reference clock \\
        OUT3 & LVDS 1.8V & $\sim$62.5038~MHz & DDMTD auxiliary clock \\
        OUT5 & LVDS 1.8V & 62.5~MHz & Auxiliary monitoring output \\
        OUT7 & LVDS 2.5V & 62.5~MHz & Auxiliary monitoring output \\
        \bottomrule
    \end{tabular}
\end{table}

\subsection{Hardware Circuit Design}

\begin{figure}[htbp]
    \centering
    \includegraphics[width=0.9\textwidth]{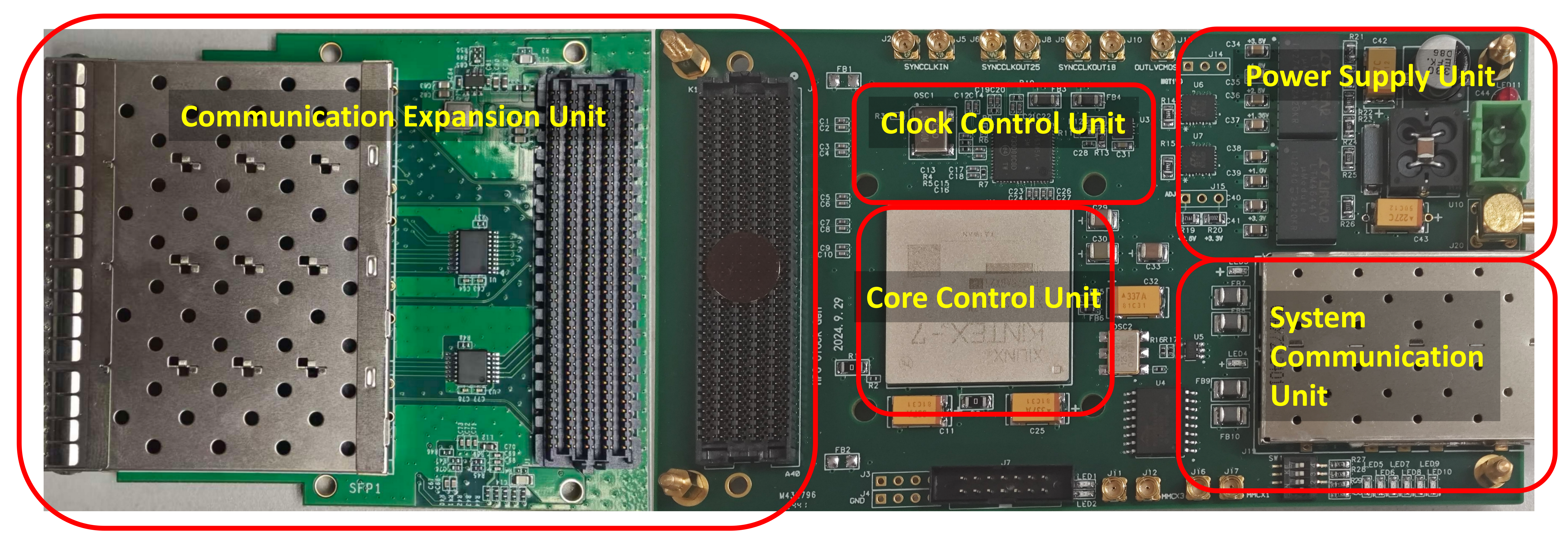}
    \caption{Photograph of the high-precision clock synchronization node hardware.}
    \label{fig:hw_photo}
\end{figure}

The clock synchronization node hardware, shown in Fig.~\ref{fig:hw_photo}, consists of four main functional units:

\textbf{Clock Control Unit}: The central clock management block, based on the Si5345A. In master mode, the Si5345A locks to an external reference or the local crystal. In slave mode, it locks to the recovered clock from the GTX receiver CDR circuit.

\textbf{Power Supply Unit}: A hybrid DCDC+LDO (low-dropout regulator) architecture provides clean power to the sensitive clock domains. Two LTM4644 quad-output step-down regulators supply the general digital circuitry, while independent low-noise LDOs provide dedicated power to the Si5345A and GTX bank analog supplies, minimizing power-supply-induced phase noise.

\textbf{System Communication Unit}: Two integrated small form-factor pluggable (SFP) interfaces plus four FMC-connected expansion SFP interfaces provide multi-port 1.25~Gbps optical connectivity. One port serves as the slave interface, while the remaining ports operate as masters distributing synchronization to downstream nodes.

\textbf{Core Control Unit}: The Xilinx Kintex-7 XC7K325T FPGA is selected as the core processing device. Compared to the Virtex-6 used in first-generation WR hardware, the Kintex-7 family offers better DDMTD performance~\cite{ref_new83}, improved GTX transceiver random jitter, and more logic resources at lower cost and power.

\subsection{New Control Loop Architecture and Noise Suppression}

\begin{figure}[htbp]
    \centering
    \begin{subfigure}[b]{0.47\textwidth}
        \includegraphics[width=\textwidth]{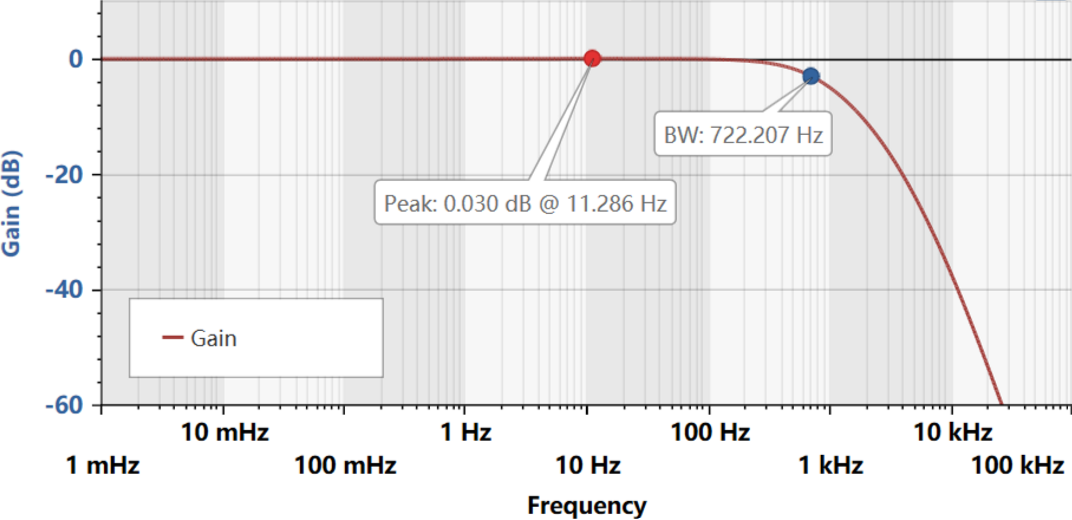}
        \caption{Closed-loop transfer function.}
    \end{subfigure}
    \begin{subfigure}[b]{0.47\textwidth}
        \includegraphics[width=\textwidth]{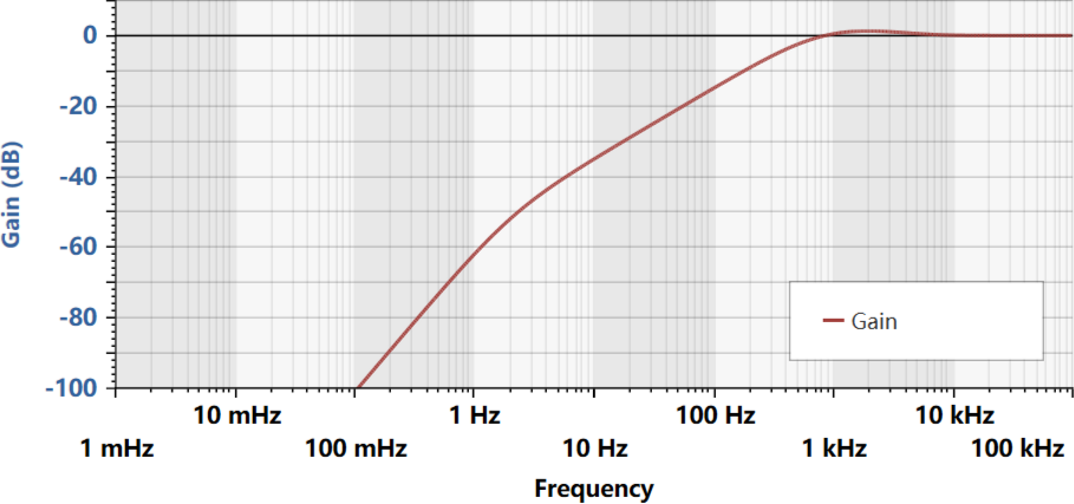}
        \caption{Error transfer function.}
    \end{subfigure}
    \caption{Bode plots of the new control loop transfer functions.}
    \label{fig:bode}
\end{figure}

In the optimized system, the WRPC (White Rabbit PTP Core) software PLL (SoftPLL) is eliminated. The fast frequency and phase locking are handled entirely by the Si5345A internal DSPLL, while the WR-PTP protocol continues to compute the master-slave offset at approximately 1~s intervals. This offset is converted into phase correction steps and applied through the DCO interface, forming a slow outer correction loop. The DDMTD now operates only in the slow outer loop, contributing phase-setpoint noise $S_{\phi,\mathrm{set}}(f)$ rather than directly modulating the actuation chain. The equivalent transfer function model gives:

\begin{equation}
    S_{\phi,\mathrm{out}}(f) = |H(f)|^2\left[S_{\phi,\mathrm{ref}}(f) + S_{\phi,\mathrm{set}}(f)\right] + |H_e(f)|^2 S_{\phi,n}(f),
\end{equation}

where $H(f)$ is the closed-loop transfer function, $H_e(f)$ is the error transfer function (high-pass) consistent with the notation of Section~\ref{sec:wr}, and \(S_{\phi,\mathrm{ref}}(f)\) denotes the equivalent phase noise power spectral density at the reference side, $S_{\phi,n}(f)$ aggregates the residual on-chip actuation noise of the Si5345A. The Bode plots of $H(f)$ and $H_e(f)$ obtained for the optimized loop are shown in Fig.~\ref{fig:bode}.

The noise suppression improvements compared to the original WR system are:

\begin{enumerate}
    \item \textbf{Actuation chain}: Removing the external DAC+VCXO eliminates the board-level analog tuning node, which is inherently sensitive to power supply ripple and ground noise coupling. The Si5345A DSPLL provides fully on-chip frequency synthesis. The custom low-noise crystal oscillator significantly reduces the reference phase noise floor.
    \item \textbf{Reference-side}: The Kintex-7 GTX offers better random jitter than Virtex-6. The Si5345A's jitter-attenuation capability reduces high-frequency noise folding. Dedicated LDO power supplies for GTX analog banks reduce low-frequency power-supply-induced jitter.
    \item \textbf{Measurement chain}: DDMTD operates outside the fast control loop, contributing only to slow phase offset computation via $S_{\phi,\mathrm{set}}(f)$. The Si5345A-generated auxiliary clock and the Kintex-7 implementation provide superior DDMTD performance.
\end{enumerate}

\section{Firmware Optimization}\label{sec:fw}
% Section 4: Firmware Optimization

\subsection{Firmware Clock Architecture}

\begin{figure}[htbp]
    \centering
    \includegraphics[width=0.75\textwidth]{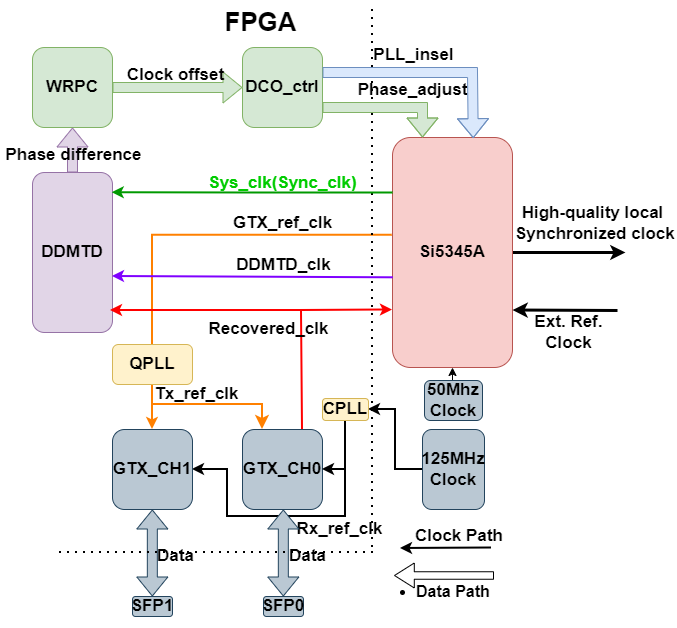}
    \caption{Block diagram of the enhanced firmware clock architecture.}
    \label{fig:fw_clock}
\end{figure}

The clock architecture of the enhanced firmware is shown in Fig.~\ref{fig:fw_clock}. After the hardware optimization, the Si5345A serves as the central clock manager. The key clock domains are:

\begin{itemize}
    \item \textbf{Recovered clock} (Recover\_clk): Generated by the GTX CH0 CDR from the incoming data stream. In slave mode, this serves as the Si5345A input reference.
    \item \textbf{Synchronized clock} (Sync\_clk): 62.5~MHz output from Si5345A OUT0, driving all time-related FPGA logic.
    \item \textbf{System clock} (Sys\_clk): Same-source as Sync\_clk, driving other functional modules.
    \item \textbf{DDMTD auxiliary clock} (DDMTD\_clk): Generated by Si5345A OUT3 at a 16385/16384 frequency ratio relative to Sync\_clk.
    \item \textbf{TX reference clock}: Si5345A OUT1, 125~MHz, for GTX transmitter serial/parallel clock generation.
    \item \textbf{RX reference clock}: Provided by an independent 125~MHz local oscillator dedicated to the GTX receiver. The RX reference must stay physically separate from the TX reference: the receiver recovers its bit clock from the incoming serial stream via the CDR, so tying the RX reference to the Si5345A would not improve receive jitter. Keeping them independent also preserves the option of separate clock-domain management within the GTX.

\end{itemize}

A critical hardware requirement is that the FPGA must support independent TX and RX reference clock sources. The Kintex-7 XC7K325T meets this requirement through its separate QPLL and CPLL paths within the GTX transceivers.

\subsection{LM32 Embedded Software Adaptation}

The LM32 embedded software running on the WRPC soft-core CPU is modified to remove the SoftPLL control modules (the Main PLL, MPLL, and the Helper PLL, HPLL). The DDMTD phase measurements are used exclusively for PTP timestamp compensation, not for fast phase control loop feedback. The master-slave offset calculated by WR-PTP is converted to DCO step values and forwarded to the hardware DCO control logic.

A 16-bit signal encodes the phase adjustment, where the MSB indicates direction (1 = delay, 0 = advance) and the lower 15 bits encode the absolute step count at 1~ps resolution.

The WR timestamp transition point calibration procedure---the standard WR routine that, at every link bring-up, scans the relative phase between the synchronized clock and the recovered clock to detect the cycle boundary at which the WR-PTP timestamp counter increments---originally depended on the MPLL to perform this scan. It is adapted for the new architecture as follows: a pre-calibration phase-fixing routine is added before scanning, ensuring that the initial phase difference between the recovered clock and the Si5345A synchronized output is fixed at a consistent state at each power-up.

\subsection{DCO Phase-Shift Control Logic}

\begin{figure}[htbp]
    \centering
    \includegraphics[width=0.6\textwidth]{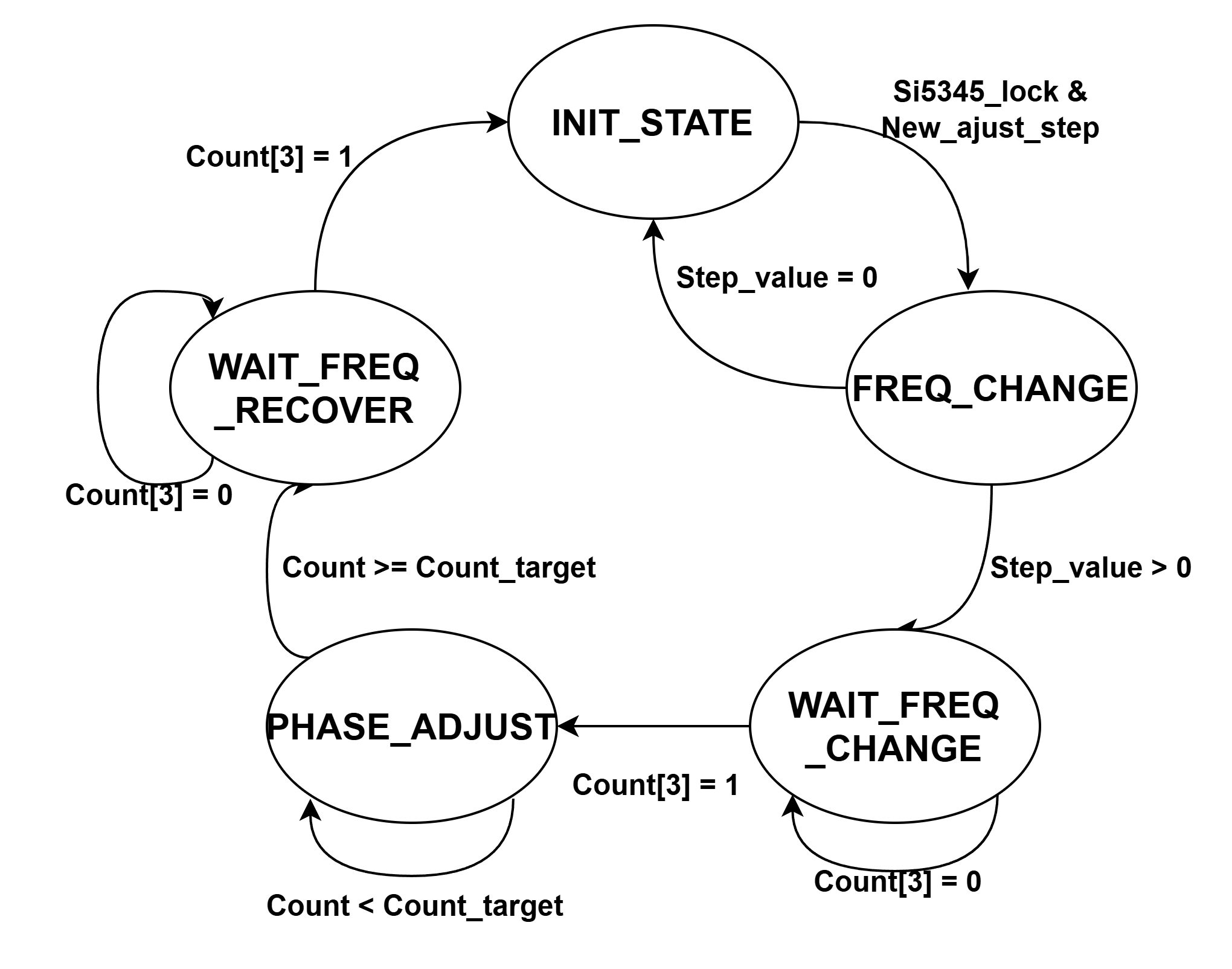}
    \caption{State transition diagram of the DCO control logic.}
    \label{fig:dco_fsm}
\end{figure}

A dedicated finite state machine (FSM), shown in Fig.~\ref{fig:dco_fsm}, translates the LM32 phase adjustment commands into Si5345A FINC/FDEC pin toggles. The FSM consists of five states:

\begin{enumerate}
    \item \textbf{INIT\_STATE}: Waits for Si5345A lock confirmation and for a new adjustment request from the LM32. When a new request arrives, the direction bit (MSB) and step count (lower 15 bits) are latched.
    \item \textbf{FREQ\_CHANGE}: Asserts FINC (delay) or FDEC (advance) according to the latched direction bit, starting the frequency offset.
    \item \textbf{WAIT\_FREQ\_CHANGE}: Holds the FINC/FDEC pin high for 8 clock cycles (128~ns at 62.5~MHz) to satisfy the Si5345A minimum 100~ns recognition window specified in the Si5345A datasheet~\cite{ref_new88}, then deasserts the pin.
    \item \textbf{PHASE\_ADJUST}: Maintains the $\pm0.5$~ppm frequency offset until an internal counter reaches $\mathit{Count}_{\mathrm{target}}$. Each step takes 125 clock cycles (2~$\mu$s) to accumulate 1~ps of phase shift, which also satisfies the Si5345A 1~$\mu$s minimum interval between consecutive FINC/FDEC toggles.
    \item \textbf{WAIT\_FREQ\_RECOVER}: Restores the nominal frequency by toggling the complementary pin for another 8 clock cycles, then returns to INIT\_STATE.
\end{enumerate}

The total phase shift is given by $\mathit{Count}_{\mathrm{target}} = \mathit{Step}_{\mathrm{value}} \times 125$, where $\mathit{Step}_{\mathrm{value}}$ is the 15-bit step count from the LM32. The additional 16 clock cycles spent in the two waiting states introduce a phase residual well below 1~ps per adjustment and are therefore neglected.

\subsection{Restart Uncertainty Optimization}

As introduced in Section~\ref{sec:wr}, the $\sim$88.8~ps peak-to-peak restart uncertainty of standard WR~\cite{ref_new87} constitutes a key bottleneck. Here we trace the effect to four independent GTX-level mechanisms and, for each, present the targeted countermeasure deployed in firmware.

\begin{figure}[htbp]
    \centering
    \includegraphics[width=0.6\textwidth]{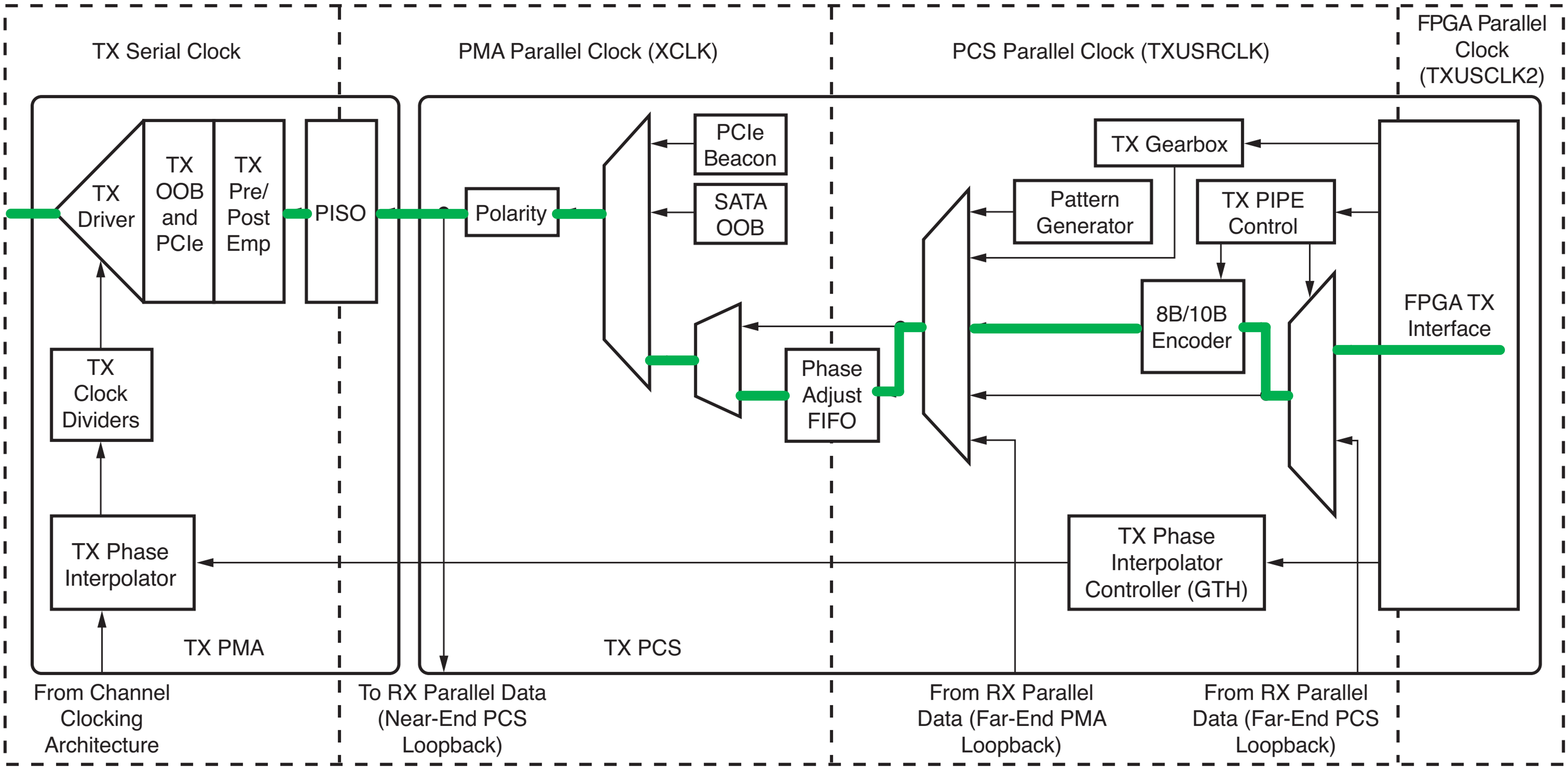}
    \caption{Schematic diagram of the GTX transmitter clock domains in the Xilinx Kintex-7 series FPGA.}
    \label{fig:gtx_rx}
\end{figure}

\textbf{(1) TXOUTCLK phase ambiguity}: The TXOUTCLK (which triggers the transmit timestamp) and the local synchronized clock (which records the timestamp value) are derived from the same 125~MHz reference but through independent division paths. Each restart can select a different phase state among $N$ possibilities, causing the TX fixed delay to change by integer clock periods. To remove this residual ambiguity, an additional DDMTD channel measures the phase between TXOUTCLK and the local synchronized clock. Since both are derived from the same Si5345A, only a finite set of discrete phase values is possible. If the measured phase does not match the preset target, the Si5345A is reset, forcing re-initialization of the GTX clock division path, and the entire initialization sequence repeats until a consistent TXOUTCLK--Sync\_clk phase is achieved.

\textbf{(2) TX Phase Adjust FIFO}: As shown in Fig.~\ref{fig:gtx_rx},The FIFO between the Physical Coding Sublayer (PCS) parallel clock domain (TXUSRCLK) and the Physical Medium Attachment (PMA) parallel clock domain (XCLK) settles at different read/write pointer offsets after each reset, introducing variable integer-cycle delays. This contribution is bypassed by enabling the GTX TX phase alignment circuit, as specified in UG476~\cite{ref_new95}: TXOUTCLK is sourced from the GTX reference clock (Si5345A OUT1, 125~MHz) and TXUSRCLK sources the TX PMA XCLK, which eliminates the parallel-in-serial-out (PISO) to XCLK phase difference.

\textbf{(3) RX byte alignment (Bit-Slide)}: During link initialization, the comma-based byte-alignment logic in the receiver (Bit-Slide in Xilinx terminology) can locks to a different byte boundary after each restart. Although WR records and compensates for the Bit-Slide value, the associated internal state change may introduce additional delay variations. This variation is eliminated by switching from automatic to manual comma alignment via the RXSLIDE interface: a custom FSM drives the alignment to a target Bit-Slide value of 0, resetting the RX if the value does not match and repeating the process until it is consistently achieved.

\textbf{(4) RX Elastic Buffer}: Similar to the TX case, the elastic buffer between the PMA and PCS clock domains in the receiver path introduces restart-dependent delay variations. It is bypassed by enabling the GTX RX phase alignment circuit, with RXOUTCLK sourced from the recovered clock and RXUSRCLK sourcing the RX XCLK.

\begin{figure}[htbp]
    \centering
    \includegraphics[width=\textwidth]{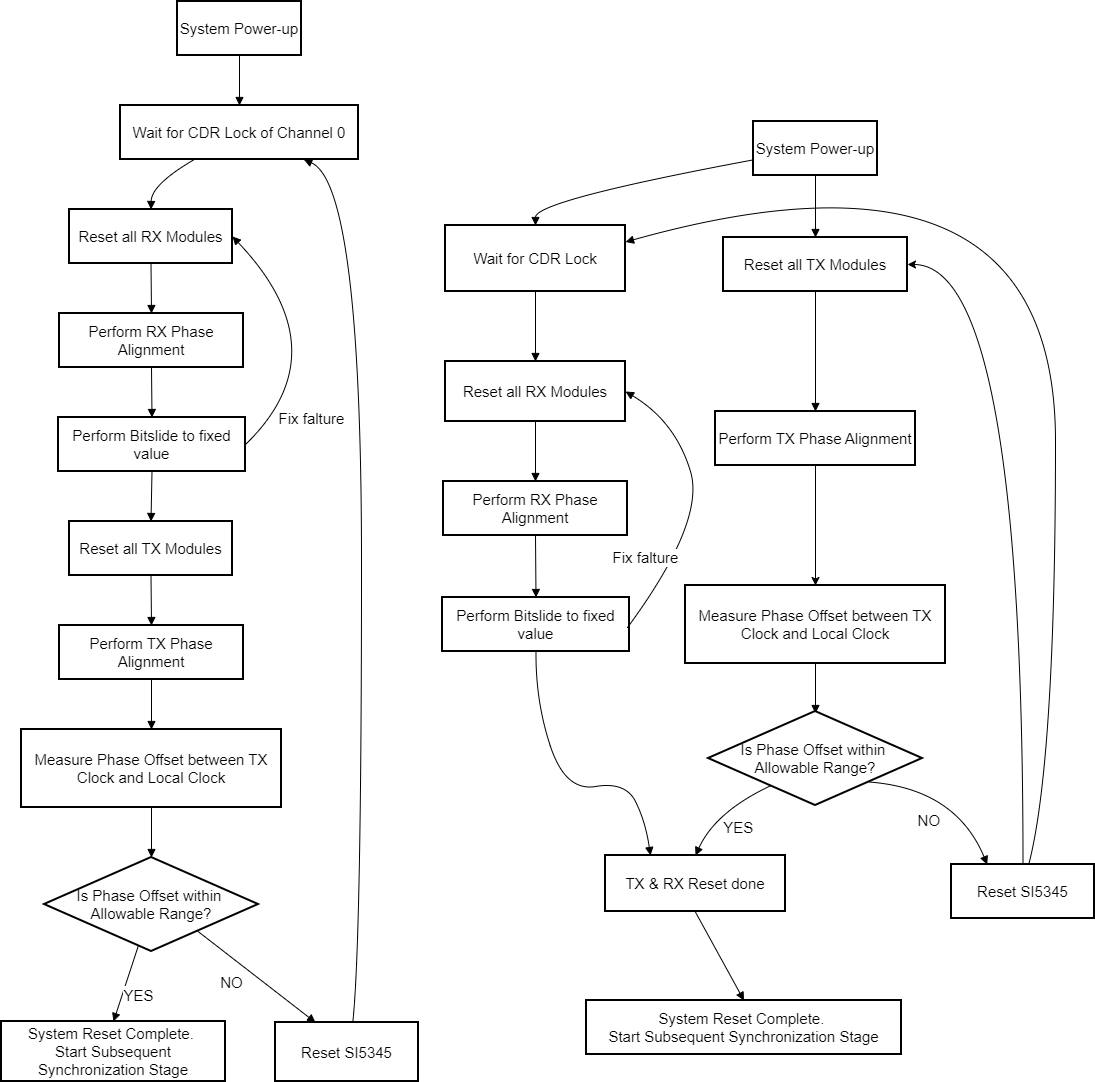}
    \caption{System initialization flowcharts after restart uncertainty optimization: (Left) slave node, (Right) master node.}
    \label{fig:init_flow}
\end{figure}

The complete initialization sequence is shown in Fig.~\ref{fig:init_flow}. The added steps introduce an average additional initialization time of less than 1 minute, which is acceptable for practical operation.

\section{Multi-Level Cascaded Global Control System}\label{sec:global}
% Section 5: Multi-Level Cascaded Global Control System

\subsection{System Architecture}

\begin{figure}[htbp]
    \centering
    \includegraphics[width=0.6\textwidth]{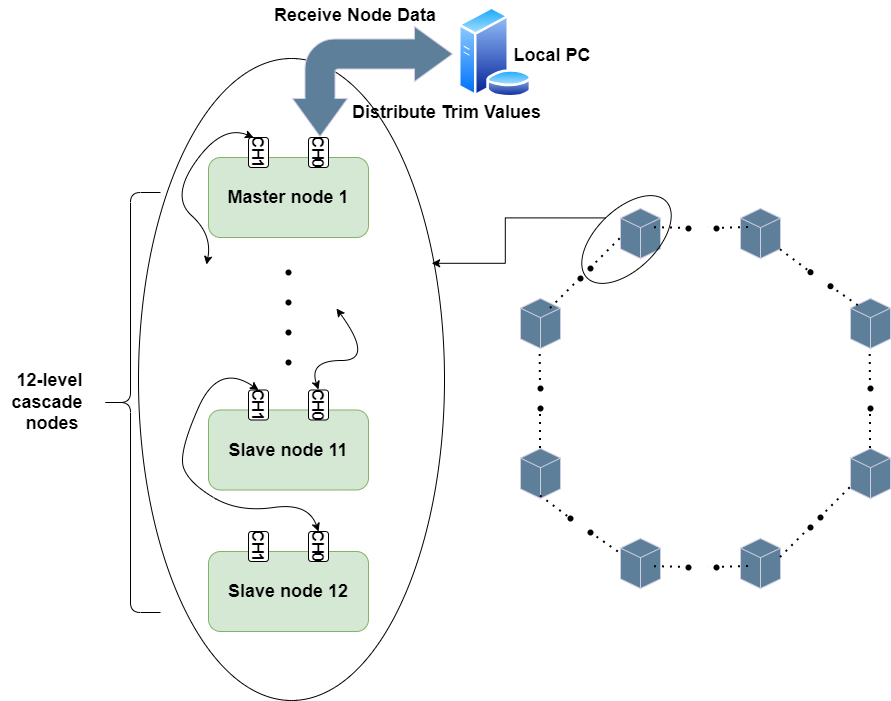}
    \caption{Deployment scheme of the multi-level cascaded global-control clock synchronization system for CEPC.}
    \label{fig:deployment}
\end{figure}

For the CEPC accelerator, where control nodes are distributed along the 100~km ring, a multi-level cascaded topology offers advantages over a pure star topology in fiber resource utilization, deployment flexibility, and maintenance efficiency. As illustrated in Fig.~\ref{fig:deployment}, the proposed architecture divides the ring into 16 cascaded subsystems, each originating from one of the eight vertical shafts and extending clockwise or counterclockwise along the tunnel. Each subsystem contains 12 cascaded nodes, covering the 192 control nodes in total.

\subsection{Motivation for Global Control}

In a multi-level cascade, each downstream node's reference signal is the synchronized output of its upstream neighbor. Consequently, upstream residual errors and control behaviors propagate and accumulate along the cascade. Under traditional ``node-autonomous'' local control, where each node independently executes a fixed-strategy PID controller, mismatches in device characteristics, power supplies, thermal conditions, and actuation dynamics can cause some nodes to over-correct while others under-correct, leading to degraded end-node synchronization precision. Furthermore, as the system scales to hundreds of nodes, manual parameter tuning becomes impractical. A global control approach that coordinates the phase compensation behavior of all nodes from a centralized PC offers significant advantages in maintainability, scalability, and the potential to incorporate advanced control strategies.

\subsection{Global Control Architecture}

\begin{figure}[htbp]
    \centering
    \includegraphics[width=0.65\textwidth]{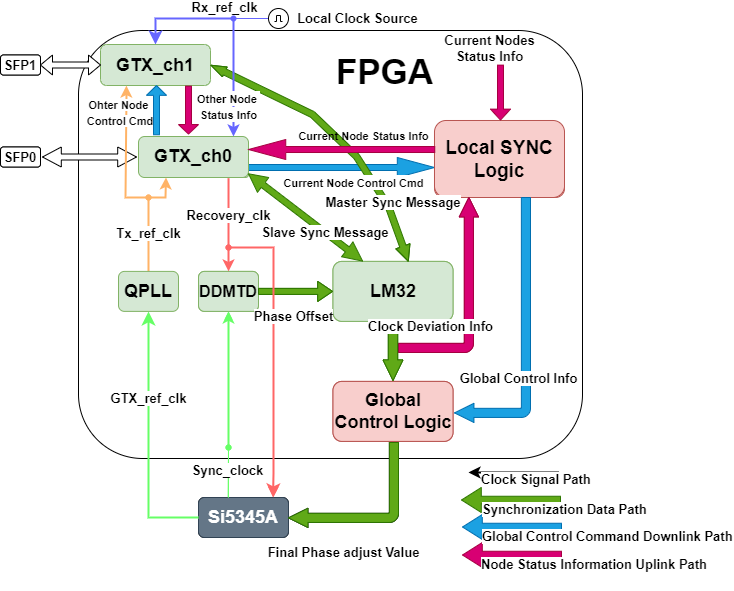}
    \caption{Block diagram of the global-control clock synchronization firmware structure.}
    \label{fig:gc_fw}
\end{figure}

The global control system adopts a ``deterministic node-side execution + centralized PC-side decision'' layered architecture, whose firmware-side block diagram is shown in Fig.~\ref{fig:gc_fw}. A two-phase operational model is used: (1) \textbf{Local synchronization phase}: each node independently achieves coarse synchronization using the local WR-PTP protocol; once the clock offset converges below a threshold, the node transitions to fine-tracking mode. (2) \textbf{Global control phase}: nodes periodically upload state information (per-hop offset, chip temperature) to the PC via Ethernet; the PC runs a multi-node coordinated control algorithm and distributes phase compensation commands back to each node.

The global control communication uses raw Ethernet MAC frames transmitted over the WR data path via the WR Streamers IP~\cite{ref_wrstreamers}, with the dual-port WRPC variant from~\cite{ref_cutewr} providing the underlying two-channel data plane. A dedicated VLAN ID distinguishes global control frames from PTP synchronization traffic. At each intermediate node, a Fabric Redirector performs address-based filtering, enabling transparent hop-by-hop forwarding along the cascaded chain. A dedicated finite state machine manages the cyclic communication with timeout and retransmission handling, while safety mechanisms include maximum step limits, validity checking on received commands, and timeout fallback to local control.

\subsubsection{Data-Link Reliability and Resource Footprint}

The global-control data path was validated on a three-node cascade (PC -- Node~0 -- Node~1 -- Node~2) under the same hop-by-hop transparent-forwarding rules used in the full deployment. A pseudo-random binary sequence (PRBS) test stream sent from Node~2 to Node~0 (VID=30, transparently forwarded by Node~1) yielded no bit errors over the test window, bounding the link bit-error rate at $< 3.79\times 10^{-12}$, and the same path sustained an effective throughput of $\sim$203~Mbps. Because the global-control traffic runs at second intervals with sub-kbps payloads, this headroom is far exceeds what the control loop demands and leaves substantial room for additional telemetry or diagnostics.

After Vivado post-implementation, the complete firmware---including the WR synchronization core, the dual-channel data plane, the node-side global-control communication, and the DCO/global control logic---occupies the resources summarized in Table~\ref{tab:res}. Logic resources are dominated by clocking and high-speed-transceiver primitives, which is consistent with the architecture; LUT, FF, and DSP utilizations remain in the single-digit percent range, leaving comfortable margin for future functional extensions.

\begin{table}[htbp]
    \centering
    \caption{Post-implementation resource utilization of the complete firmware on Xilinx XC7K325T.}
    \label{tab:res}
    \begin{tabular}{lc}
        \toprule
        Resource & Utilization (\%) \\
        \midrule
        LUT     & 7.51 \\
        LUTRAM  & 0.19 \\
        FF      & 4.42 \\
        BRAM    & 15.51 \\
        DSP     & 0.71 \\
        IO      & 9.75 \\
        GT      & 25.00 \\
        BUFG    & 53.13 \\
        MMCM    & 40.00 \\
        \bottomrule
    \end{tabular}
\end{table}

\subsection{PC-Side Controller with RL-Based Parameter Auto-Tuning}

\subsubsection{Control Structure}

The PC-side controller is organized as an ``error feedback + temperature feed-forward'' structure. For each slave node $i$, the proposed controller takes a PID-like form
\begin{equation}
    u_i(k) = K_p^{(i)} e_i(k) + K_i^{(i)} \sum_{j=0}^{k} e_i(j) + K_d^{(i)} \bigl(e_i(k) - e_i(k-1)\bigr) + f_i\bigl(T_i(k)\bigr),
\end{equation}
where $e_i(k)$ is the temperature-corrected per-hop clock offset, $\{K_p^{(i)},K_i^{(i)},K_d^{(i)}\}$ are the PID gains, and $f_i(T_i(k))$ is a linear temperature-dependent feed-forward term whose slope is calibrated from a dedicated warm/heat experiment (see Section~\ref{subsec:tempcomp}). For purely PI operation, the differential term is fixed to zero. The control output is amplitude-limited to $\pm30$~ps per cycle and quantized to the 1~ps DCO step before being sent to the corresponding node.

\subsubsection{Temperature Feed-Forward Compensation}\label{subsec:tempcomp}

\begin{figure}[htbp]
    \centering
    \begin{subfigure}[b]{0.47\textwidth}
        \includegraphics[width=\textwidth]{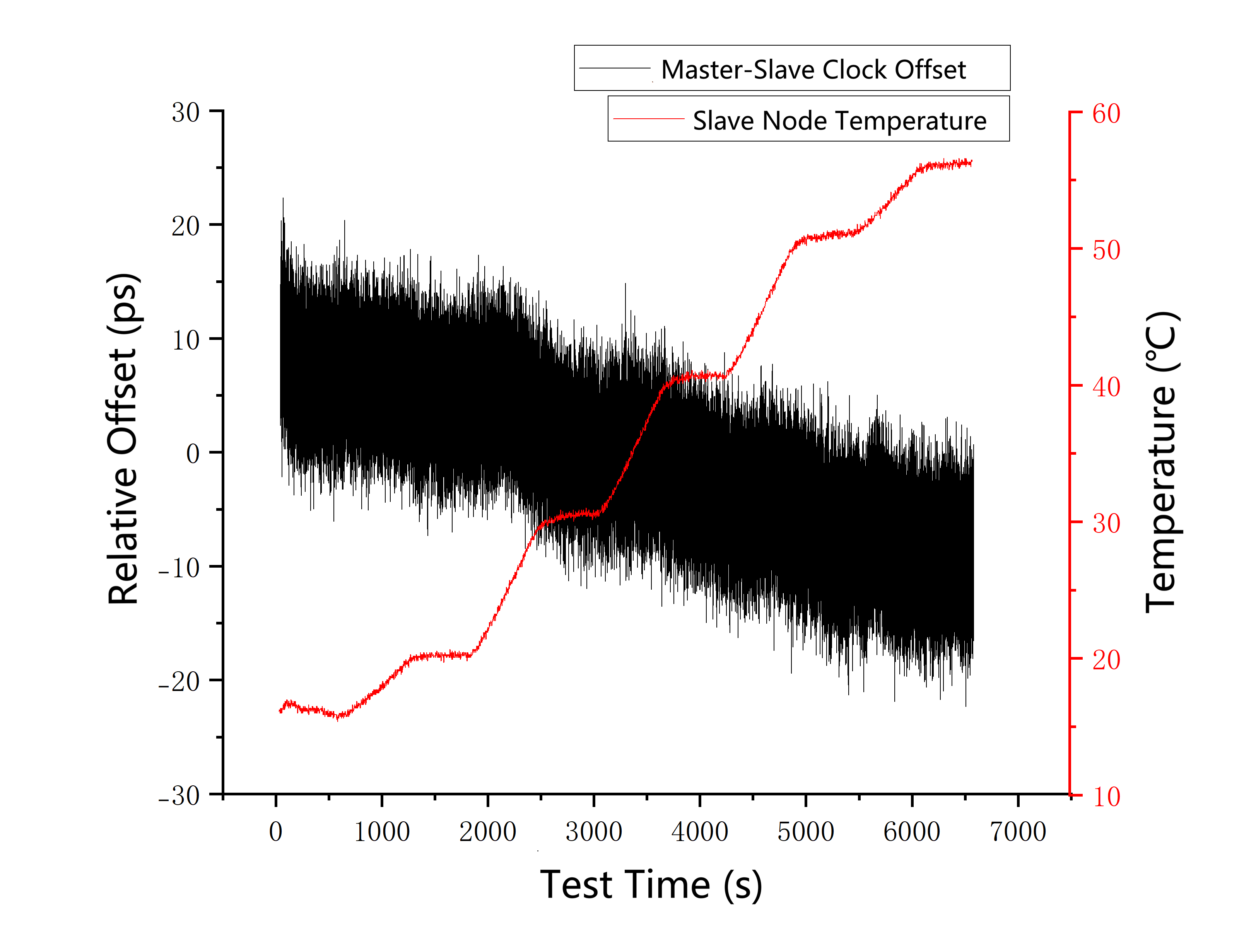}
        \caption{Before compensation}
        \label{fig:temp_before}
    \end{subfigure}
    \begin{subfigure}[b]{0.47\textwidth}
        \includegraphics[width=\textwidth]{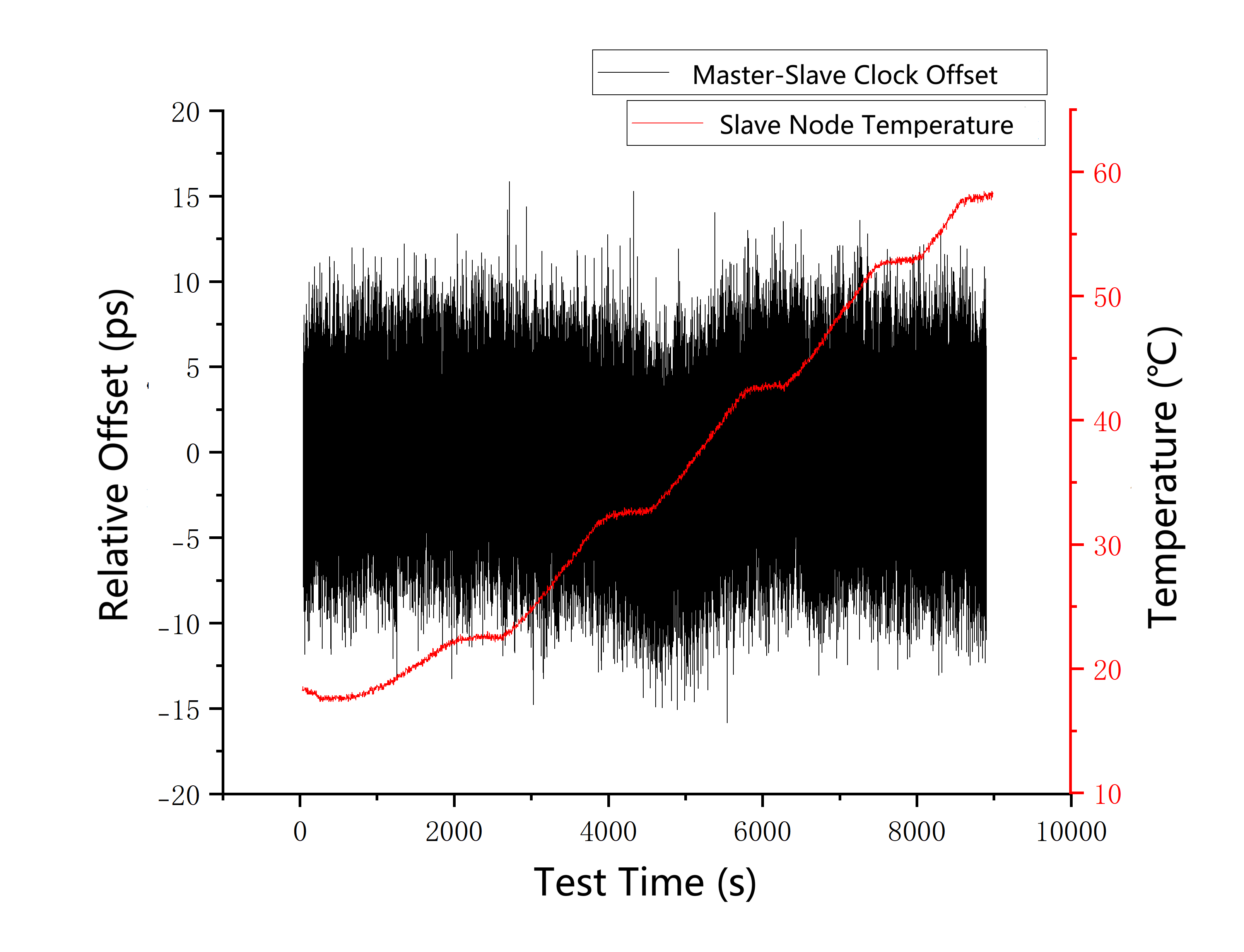}
        \caption{After compensation}
        \label{fig:temp_after}
    \end{subfigure}
    \caption{Temperature compensation calibration results.}
    \label{fig:temp_comp}
\end{figure}

Node temperature is sampled by the on-chip Xilinx XADC (Xilinx Analog-to-Digital Converter) and reported together with the local offset in every control cycle. The temperature coefficient was calibrated on a point-to-point link using two scenarios:
(i) the master held at constant temperature while the slave was warmed from $15\,^\circ$C to $55\,^\circ$C;
(ii) master and slave warmed together over the same range. Scenario~(i) produced an offset change of $\sim$30~ps with a linear slope of $-0.7578\,\mathrm{ps}/^\circ$C, while scenario~(ii) produced no significant change, confirming that the dominant drift is driven by inter-node temperature \emph{difference} rather than absolute temperature. A linear feed-forward model with this slope was implemented in the PC controller; after compensation, the residual coefficient under scenario~(i) drops to $-0.00425\,\mathrm{ps}/^\circ$C, even with a $40\,^\circ$C inter-node temperature difference. The before/after calibration curves are shown in Fig.~\ref{fig:temp_comp}.

\subsubsection{RL-Based Parameter Auto-Tuning with TD3}

The global control loop runs at a roughly 1-second intervals; its objective is to drive multi-node clock offsets to convergence under finite phase-step resolution (1~ps) and update-period constraints. Manual tuning becomes impractical as the number of nodes grows, so PID parameter optimization is cast as a reinforcement learning (RL) problem, for which we employ the TD3 (Twin-Delayed Deep Deterministic Policy Gradient) algorithm~\cite{ref_td3}, with the structure summarized below.

\paragraph{State, action, and reward.} For an $N$-node cascade, the state $s(k)$ stacks $\{e_i(k), I_i(k), D_i(k)\}_{i=1}^{N}$ with $I_i(k)=\sum_{t=0}^{k} e_i(t)$ and $D_i(k) = e_i(k)-e_i(k-1)$, yielding a $3N$-dimensional input. The action $u(k)\in \mathbb{R}^{N}$ contains the per-cycle phase compensation in ps for each slave node; it is clipped to $\pm30$~ps and integer-quantized before being dispatched. The reward function combines a time penalty, per-node and global improvement terms, and a steady-state bonus:
\begin{equation}
    r(k) = -c_0 - \lambda_1 \sum_{i=1}^{N}\!\bigl(|e_i(k)|-|e_i(k-1)|\bigr) - \lambda_2 \bigl(E(k)-E(k-1)\bigr) + r_{\mathrm{bonus}}(k),
\end{equation}
where $E(k)=\sum_i |e_i(k)|$ and $r_{\mathrm{bonus}}(k)=R_s$ when $E(k)\leq \varepsilon$, otherwise 0. An episode terminates with success when $E(k)\leq\varepsilon$ for several consecutive cycles, or with failure when $E(k)\geq\varepsilon_{\mathrm{hi}}$ persists.

\paragraph{Linear Actor as PID surrogate.} Rather than running the RL policy online, we restrict the Actor to a single fully-connected linear layer with no activation:
\begin{equation}
    u(k) = W\,s(k) + b.
\end{equation}
Because $s(k)$ already contains the proportional, integral, and differential terms for every node, a linear mapping on $s(k)$ is mathematically expressive enough to cover the entire classical multi-node PID solution space: the block-diagonal entries of $W\in\mathbb{R}^{N\times3N}$ correspond exactly to the per-node gains $\{K_p^{(i)},K_i^{(i)},K_d^{(i)}\}$ in Eq.~(5.1), while the off-diagonal blocks capture the inter-node coupling that a node-autonomous PID cannot represent. The bias $b\in\mathbb{R}^N$ absorbs any residual steady-state offset. Non-linearity is thus kept entirely within the Critic, whose only job is to approximate the long-horizon return during training. At deployment time, the Critic is discarded and the controller is a deterministic linear PID. This preserves low computational cost, full determinism, and amenability to classical stability analysis online, while letting the training stage handle the multi-node, multi-parameter optimization.

\paragraph{TD3 configuration.} The Actor is paired with two Critic networks, each implemented as ``state and action embedding $\rightarrow$ concatenation $\rightarrow$ two fully-connected hidden layers $\rightarrow$ scalar $Q$ output''. We use a discount factor $\gamma=0.99$, soft target update $\tau=5\times10^{-3}$, replay buffer of $10^6$ transitions, batch size 256, and policy delay of 2. Exploration noise is added during pre-training, and clipped target-policy smoothing noise is injected when computing TD targets to suppress $Q$ overestimation.

\paragraph{Two-stage training and parameter transfer.} Training proceeds in two stages to avoid the cost and risk of starting from scratch on the hardware. (i) In a virtual environment with the simplified dynamics $e(k+1) = e(k) - u(k) + n(k)$, where $n(k)$ is zero-mean Gaussian noise, the agent rapidly learns a viable PID surrogate for a single node. (ii) The pretrained weights are loaded onto the real cascaded hardware as initial values, and TD3 continues fine-tuning under genuine noise, quantization, and communication jitter. For the multi-node cascade, the single-node weights are block-replicated along the diagonal of $W$ to provide a sensible warm start; on-system fine-tuning is then performed on a 7-level cascade. Reward curves converge within a few hundred episodes in both the virtual and real-system stages, and the resulting controller reduces multi-node offsets to near zero within a few control cycles, Fig.~\ref{fig:temp_comp} shows reward curves at different training stages and synchronized clock offset trajectories of the test nodes. The current global controller runs on a standard PC connected to the cascade master node via Ethernet; migration to an FPGA-based embedded platform is planned for enhanced determinism in future deployment.

\begin{figure}[!htbp]
    \centering
    \begin{subfigure}[b]{0.32\textwidth}
        \centering
        \includegraphics[width=\textwidth]{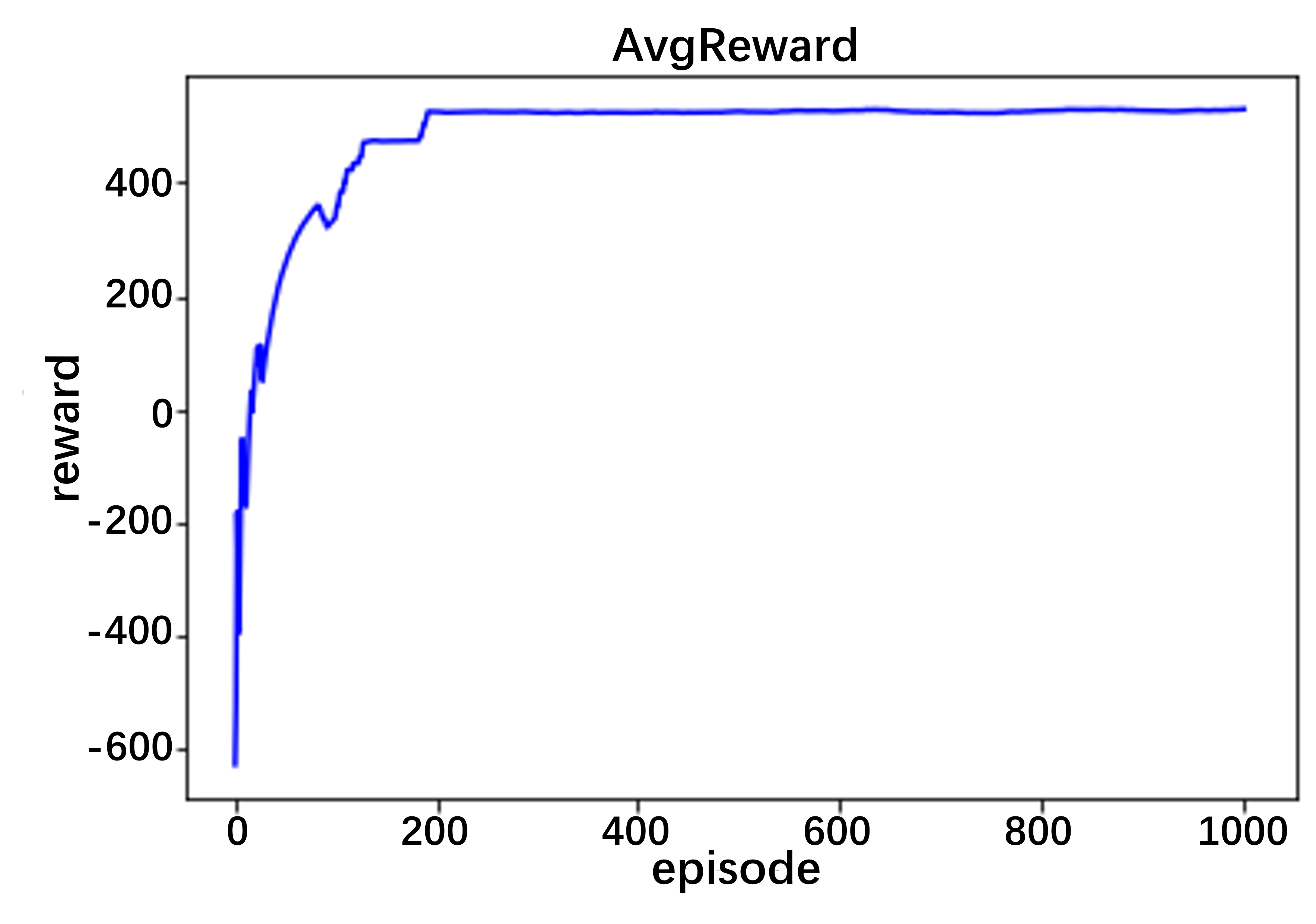}
        \caption{}
        \label{image59a}
    \end{subfigure}
    \begin{subfigure}[b]{0.32\textwidth}
        \centering
        \includegraphics[width=\textwidth]{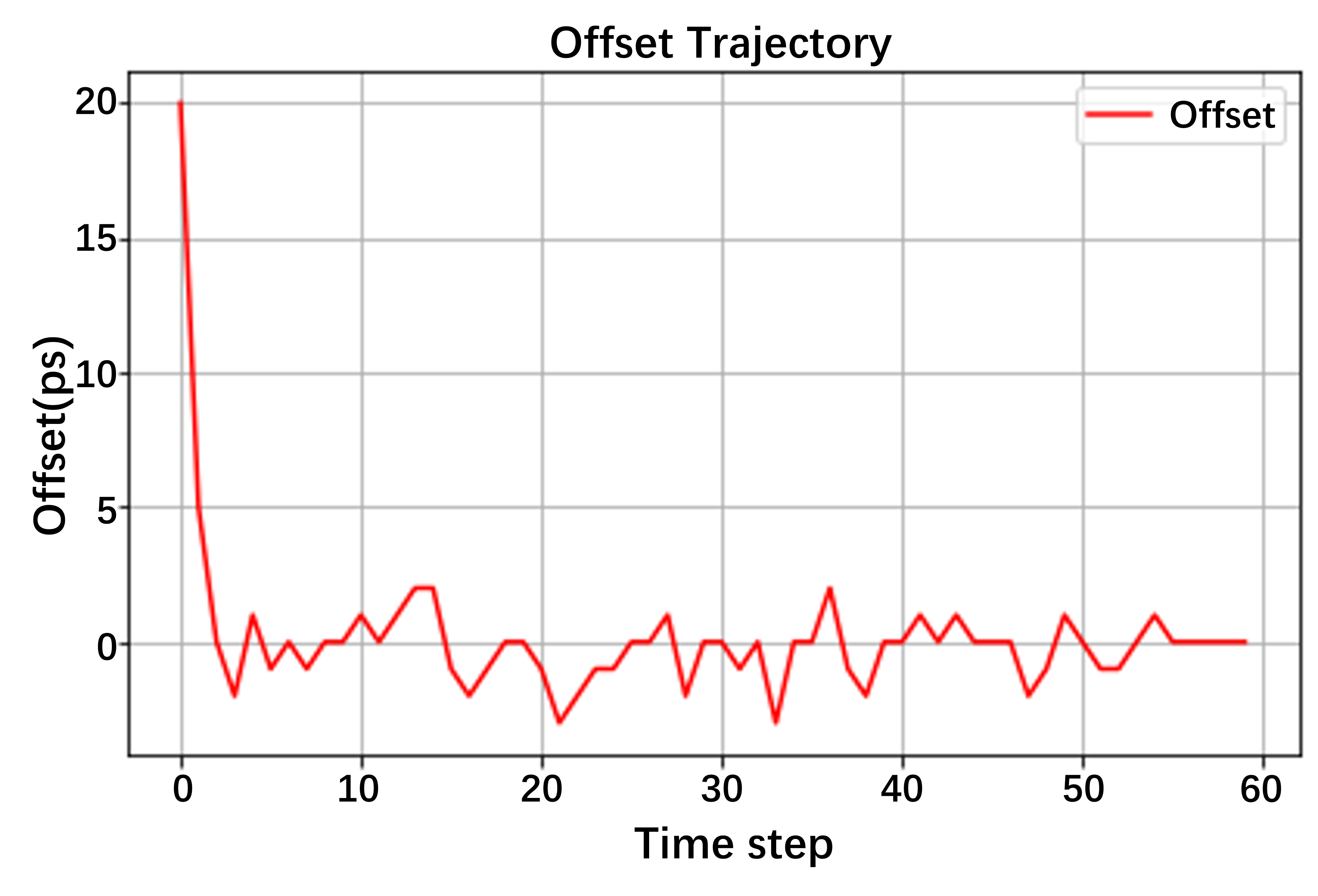}
        \caption{}
        \label{image59b}
    \end{subfigure}
    \begin{subfigure}[b]{0.32\textwidth}
        \centering
        \includegraphics[width=\textwidth]{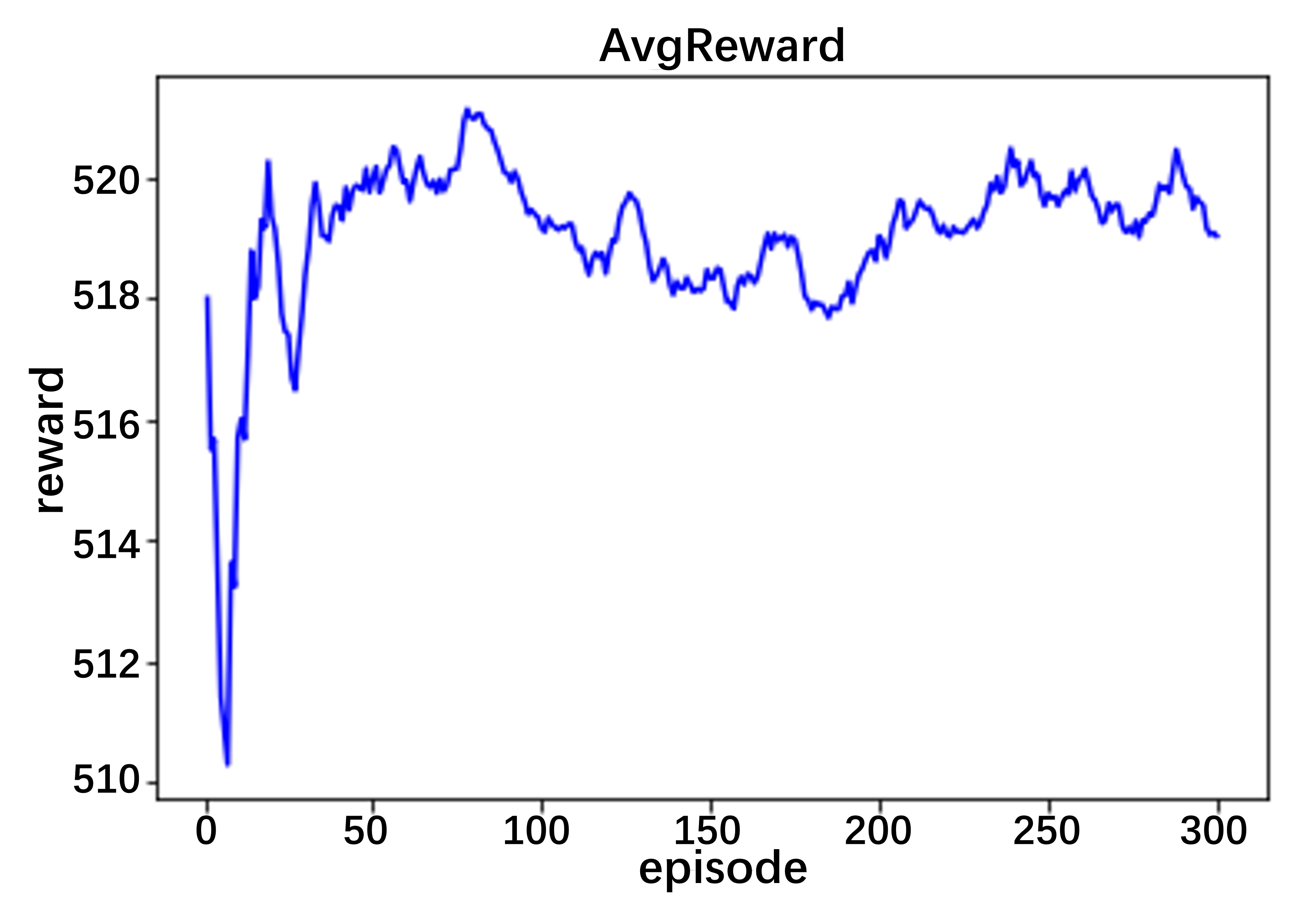}
        \caption{}
        \label{image59c}
    \end{subfigure}

    \begin{subfigure}[b]{0.32\textwidth}
        \centering
        \includegraphics[width=\textwidth]{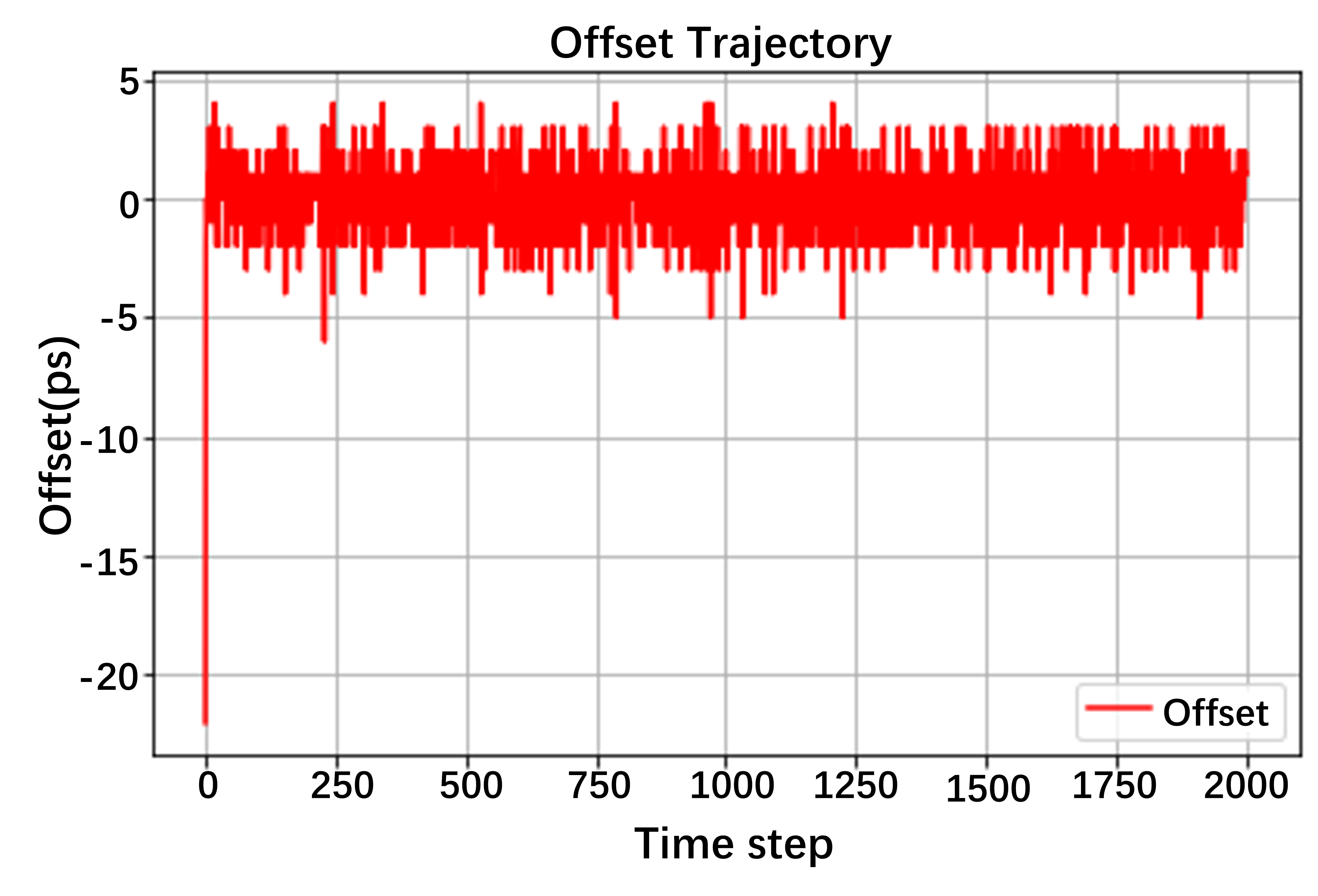}
        \caption{}
        \label{image59d}
    \end{subfigure}
    \begin{subfigure}[b]{0.32\textwidth}
        \centering
        \includegraphics[width=\textwidth]{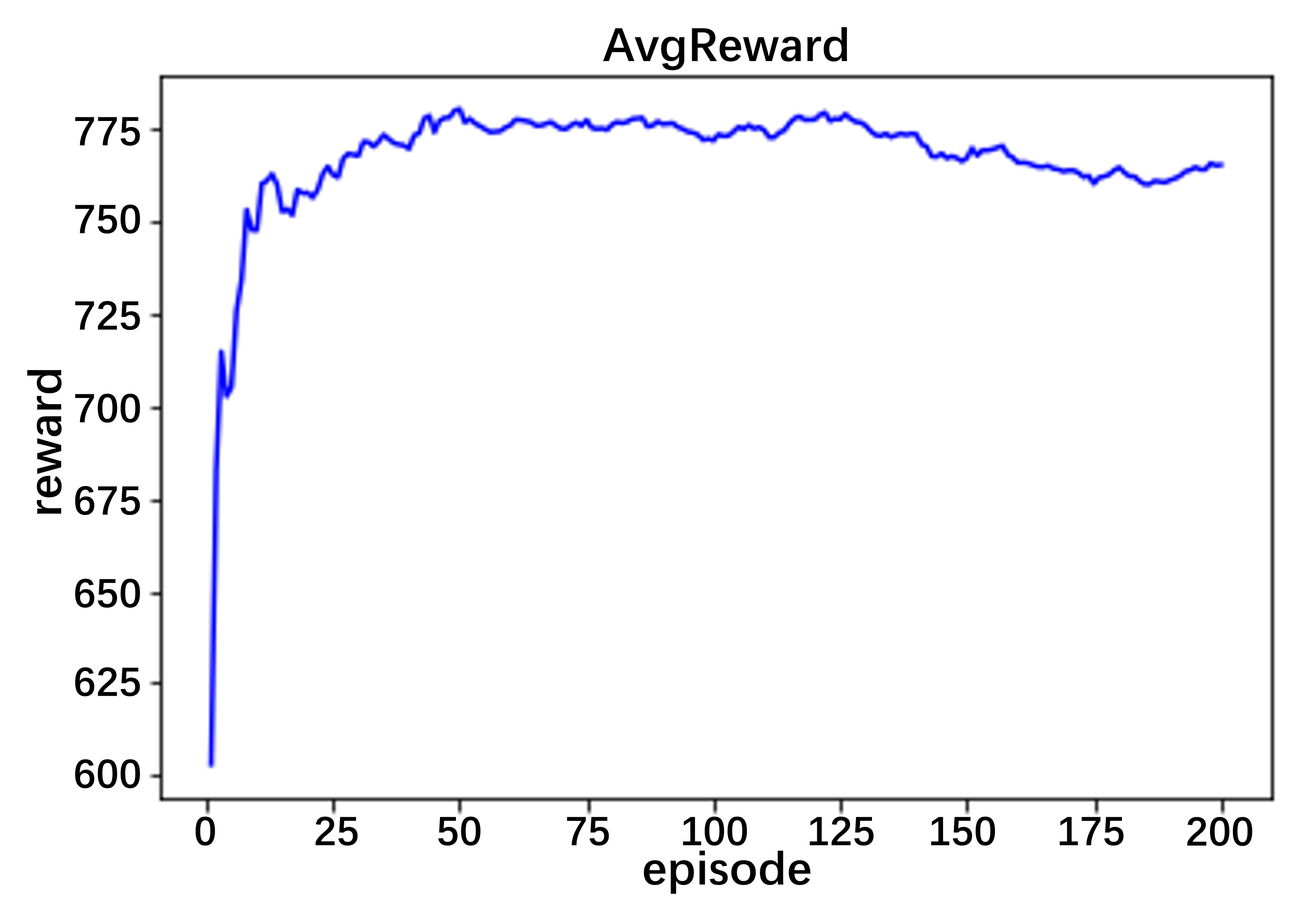}
        \caption{}
        \label{image59e}
    \end{subfigure}
    \begin{subfigure}[b]{0.32\textwidth}
        \centering
        \includegraphics[width=\textwidth]{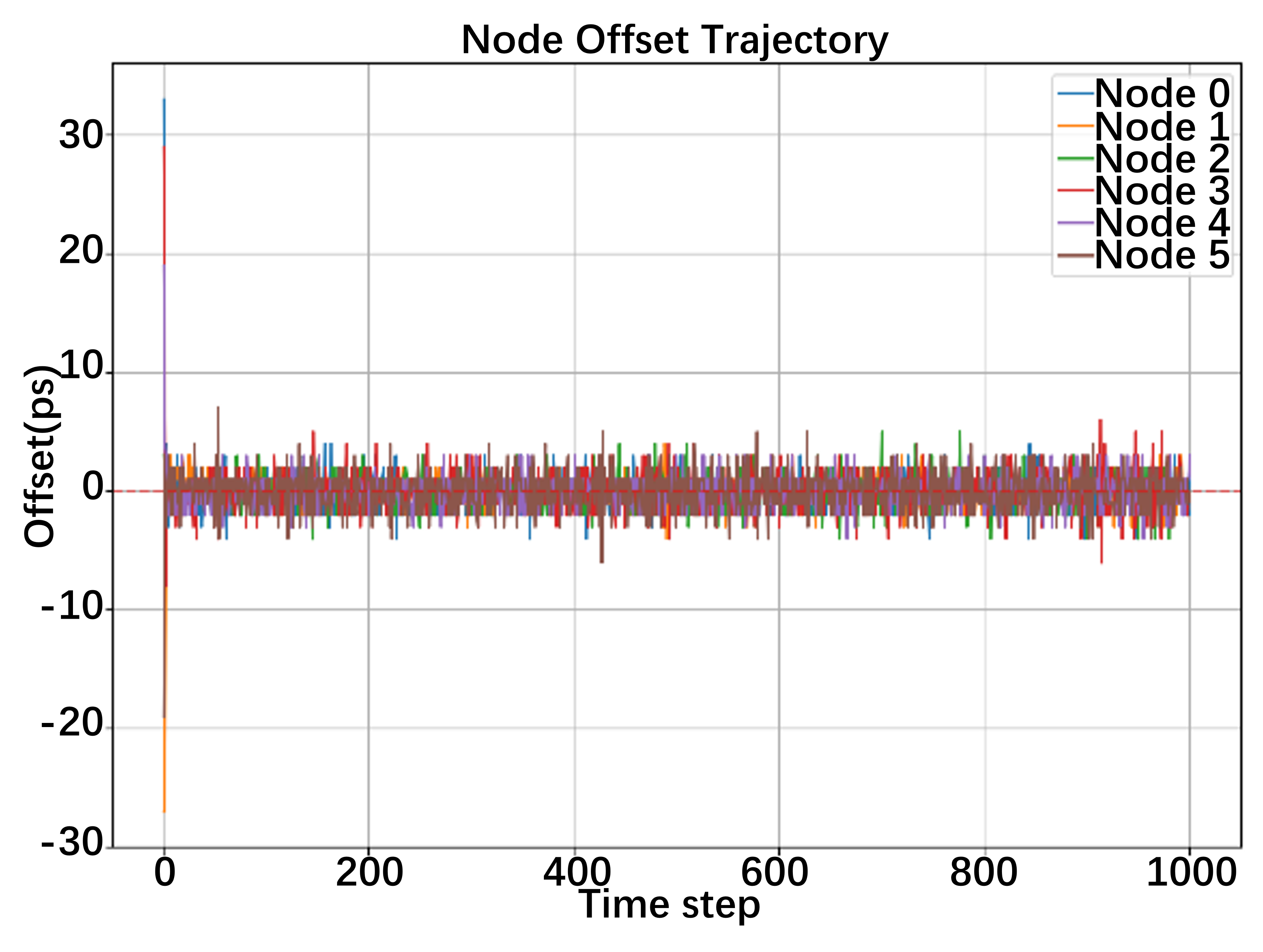}
        \caption{}
        \label{image59f}
    \end{subfigure}
    \caption{ Reward curves at different training stages and synchronized clock offset trajectories of the test nodes.
(a) Pre-training reward curve in virtual environment.
(b) Evolution trajectory of node clock offset during tests in virtual environment.
(c) Online training reward curve under single-node configuration.
(d) Evolution trajectory of node clock offset during tests for single-node setup.
(e) Online training reward curve under multi-node configuration.
(f) Evolution trajectory of node clock offset during tests for multi-node setup.}
    \label{image59}
\end{figure}

\section{Experimental Results}\label{sec:results}
% Section 6: Experimental Results

\subsection{Test Setup}

All tests were conducted with the clock synchronization nodes described in Section~\ref{sec:hw}. Synchronized clock signals are extracted via MMCX connectors, converted from LVDS to single-ended using 3~GHz bandwidth balun modules, and measured with either a Tektronix DPO~7254C (2.5~GHz, 40~GS/s, denoted scope~A) or a Teledyne LeCroy SDA~820Zi-B (20~GHz, 80~GS/s, denoted scope~B). All systems are pre-warmed for 30~min before data acquisition; the long-term run lasts 25~h, and all other long-time tests run for at least 8~h. \emph{Synchronization precision} is the standard deviation of the slave-vs-master clock skew over the measurement window. Where reported, ``peak-to-peak'' refers to the same window. For PPS-offset (accuracy) measurements, the system is first calibrated using the standard WR fiber-asymmetry coefficient $\alpha$ and the fixed TX/RX hardware delays, both extracted from a three-fiber-length measurement of $\mathrm{delay}_{MM}$ and a swap-and-average oscilloscope comparison; the calibration procedure follows the standard WR practice and is not the subject of this paper.

The standard WR baseline used for comparison is taken from~\cite{ref_new87}: point-to-point precision $\sim$20--30~ps, restart uncertainty of 14.7~ps (std.\ dev.) and 88.8~ps (peak-to-peak), and synchronized-clock jitter on the order of tens of picoseconds. The complete test matrix is summarized in Table~\ref{tab:test_matrix}. Different cascade depths and durations are used in different tests due to oscilloscope channel count and node availability constraints: the 12-level cascade exhausts the available probing channels of scope~B for a single snapshot, so the long-term continuous run and the cycle-by-cycle jitter characterization were performed on a smaller 4-level cascade that could remain fully instrumented for the entire test window.

\begin{table}[htbp]
    \centering
    \caption{Test matrix of the experimental campaign.}
    \label{tab:test_matrix}
    \begin{tabular}{lcccc}
        \toprule
        Test & Cascade depth & Fiber per hop & Duration & Scope \\
        \midrule
        Point-to-point precision        & 2 nodes (1 hop)  & 1~m / 30~km / 50~km & $>$8~h & A \\
        Cascade, constant temperature   & 12 nodes         & 1~m                 & $>$8~h & B \\
        Cascade, variable temperature   & 12 nodes         & 1~m                 & $\sim$1~h profile & B \\
        Long-term stability             & 4 nodes          & 1~m                 & 25~h   & B \\
        Synchronized-clock jitter (TIE) & 4 nodes          & 1~m                 & $>$8~h & B \\
        Restart uncertainty             & 2 nodes (1 hop)  & 1~m                 & 60 cycles & B \\
        \bottomrule
    \end{tabular}
\end{table}

\subsection{Point-to-Point Synchronization Performance}

\begin{figure}[htbp]
    \centering
    \begin{subfigure}[b]{0.32\textwidth}
        \includegraphics[width=\textwidth]{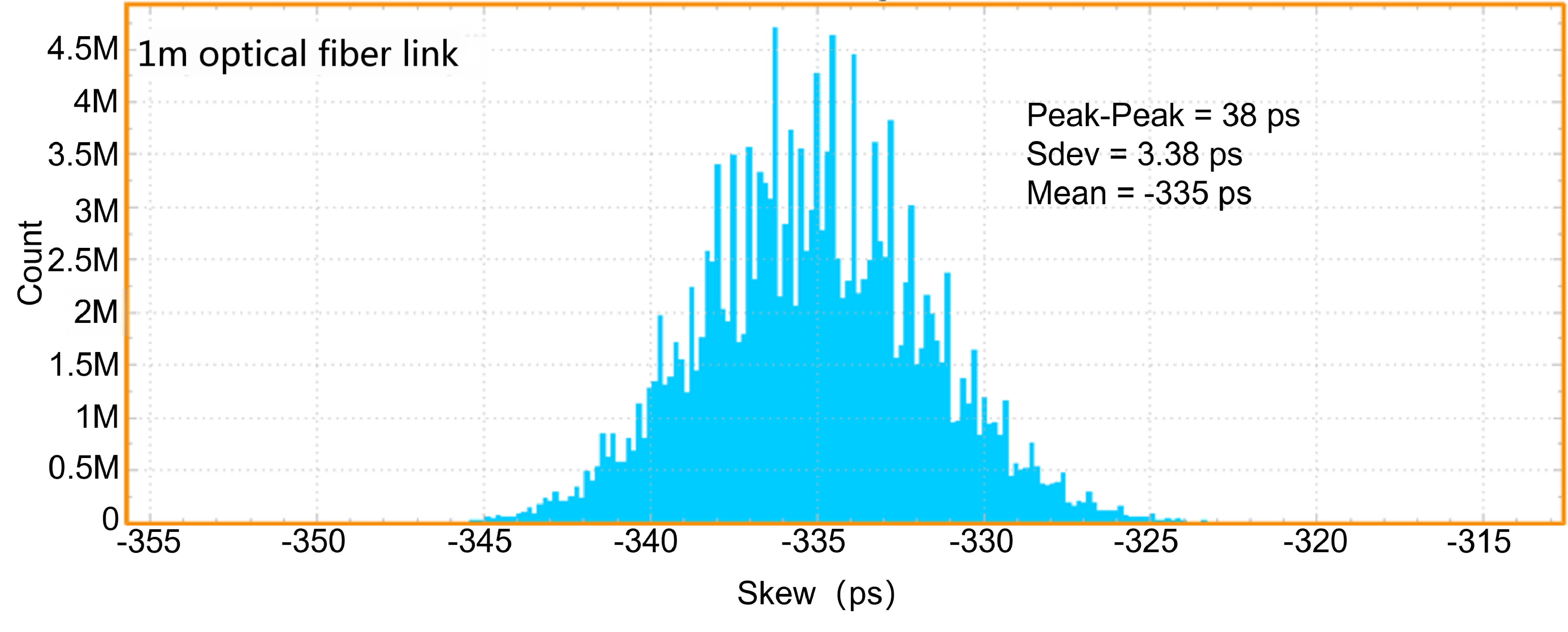}
        \caption{1~m fiber, $\sigma=3.38$~ps}
    \end{subfigure}
    \begin{subfigure}[b]{0.32\textwidth}
        \includegraphics[width=\textwidth]{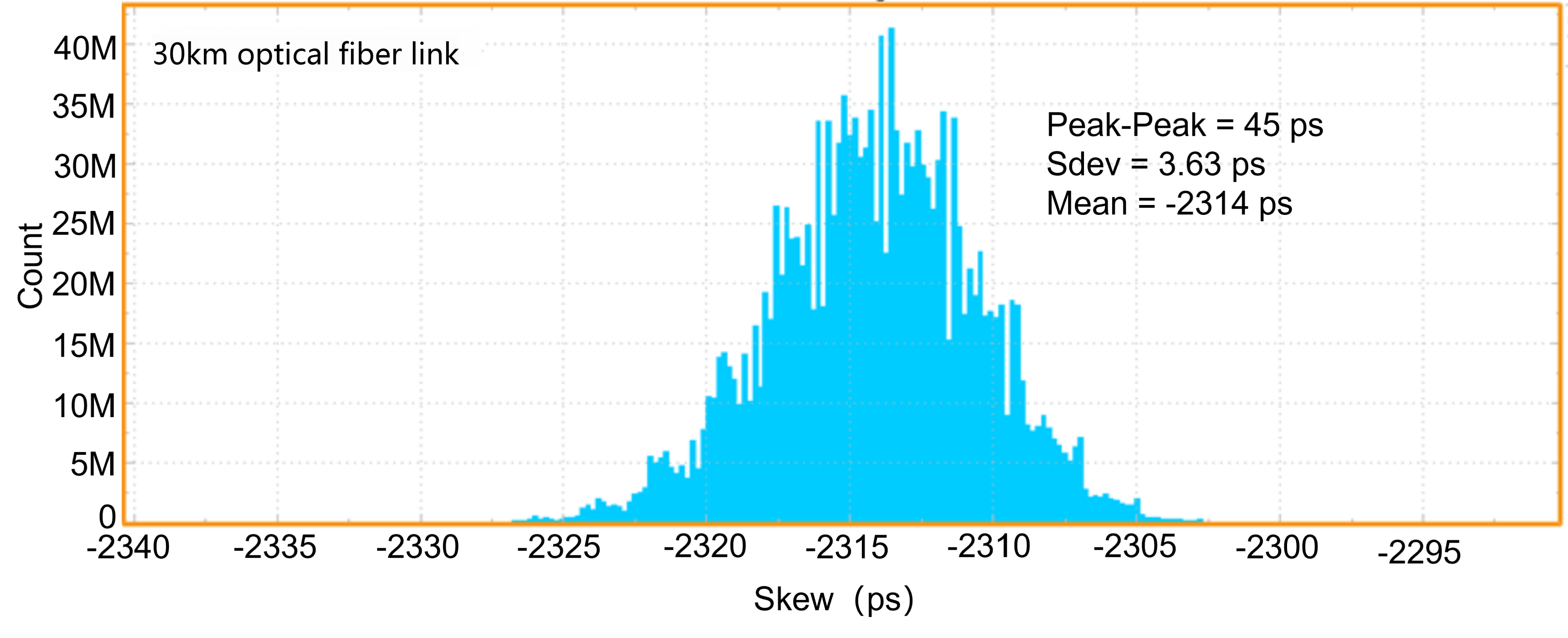}
        \caption{30~km fiber, $\sigma=3.63$~ps}
    \end{subfigure}
    \begin{subfigure}[b]{0.32\textwidth}
        \includegraphics[width=\textwidth]{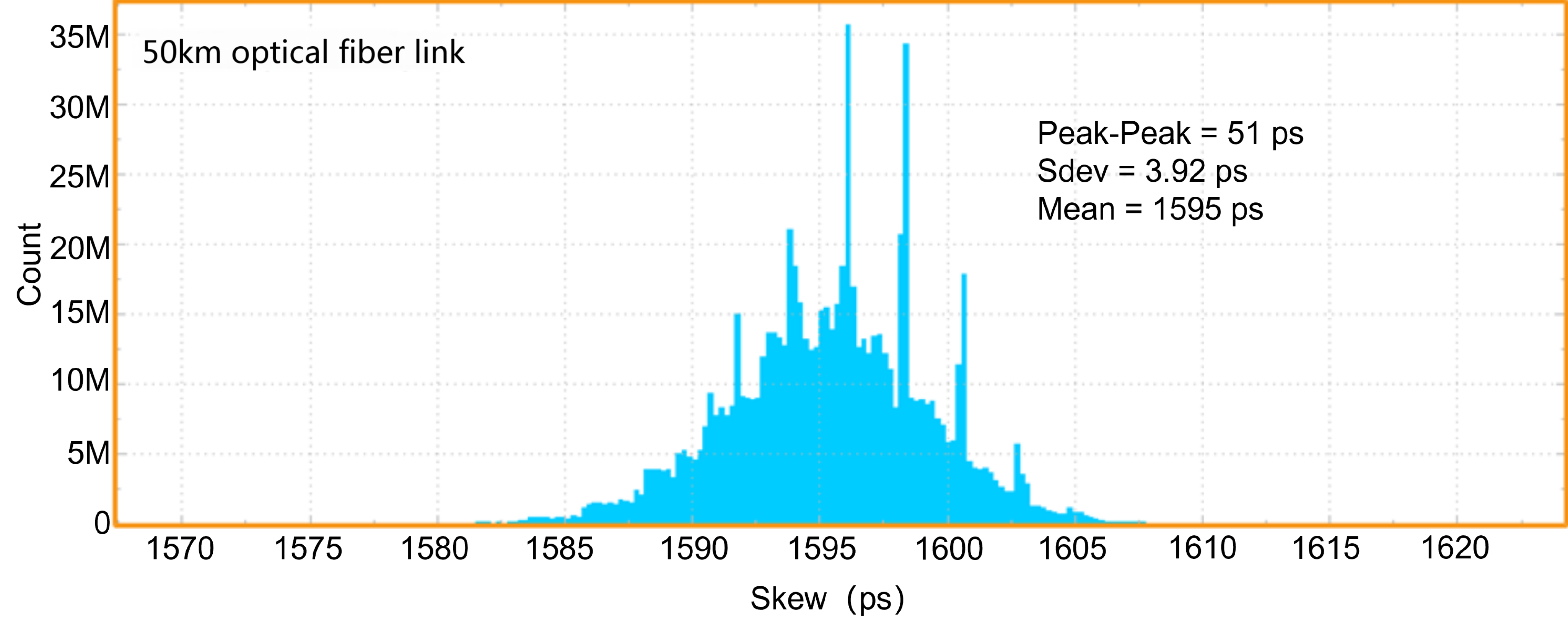}
        \caption{50~km fiber, $\sigma=3.92$~ps}
    \end{subfigure}
    \caption{Point-to-point clock synchronization measured with scope~A over a $>$8~h window. The 50~km case bounds the maximum link length expected in the CEPC tree topology.}
    \label{fig:p2p_results}
\end{figure}

Figure~\ref{fig:p2p_results} shows the point-to-point test results. The synchronization precision is 3.38~ps at 1~m, 3.63~ps at 30~km, and 3.92~ps at 50~km. The 1~m case isolates the intrinsic performance of the node electronics, the 30~km case is representative of mid-range inter-cavern spans in the CEPC ring, and the 50~km case bounds the worst-case link length. The precision degrades by only $\sim$0.5~ps between 1~m and 50~km, confirming that the dominant noise contribution is local to the node rather than fiber-link related. Even at 50~km the precision remains below 4~ps, a $\sim$6$\times$ improvement over the standard WR baseline cited above. This also verifies that the synchronization accuracy of the system fully meets the requirements even with a star topology in CEPC accelerator.

\subsection{Multi-Level Cascaded Global Control Performance}

A 12-level cascaded system was constructed using 12 clock synchronization nodes interconnected with 1~m short fibers, with the PC-side global controller (TD3-tuned PID + temperature feed-forward) active throughout. Levels~5, 9 and~12 were probed with scope~B.

\begin{figure}[htbp]
    \centering
    \begin{subfigure}[b]{0.32\textwidth}
        \includegraphics[width=\textwidth]{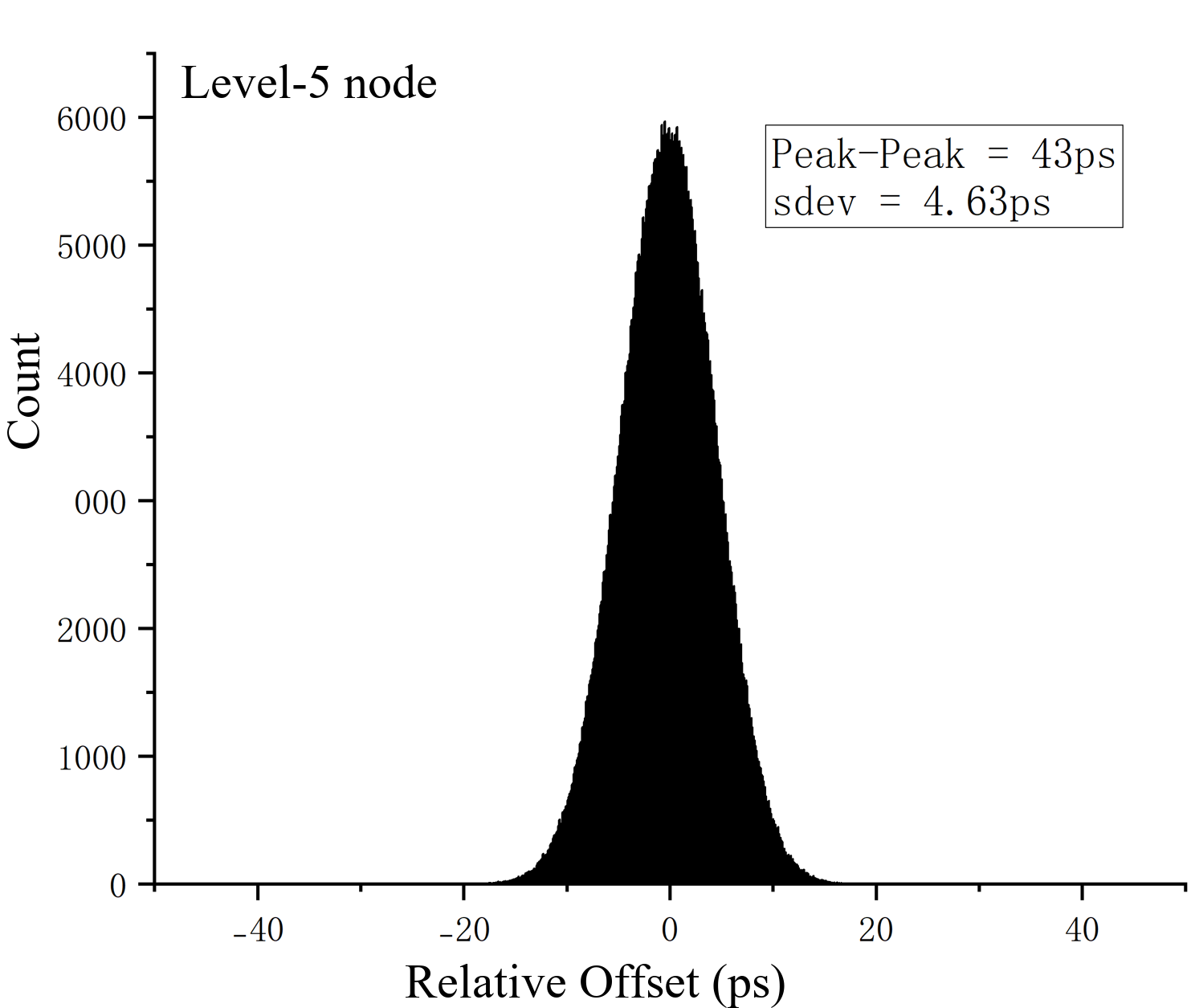}
        \caption{Level~5, $\sigma=4.63$~ps}
    \end{subfigure}
    \begin{subfigure}[b]{0.32\textwidth}
        \includegraphics[width=\textwidth]{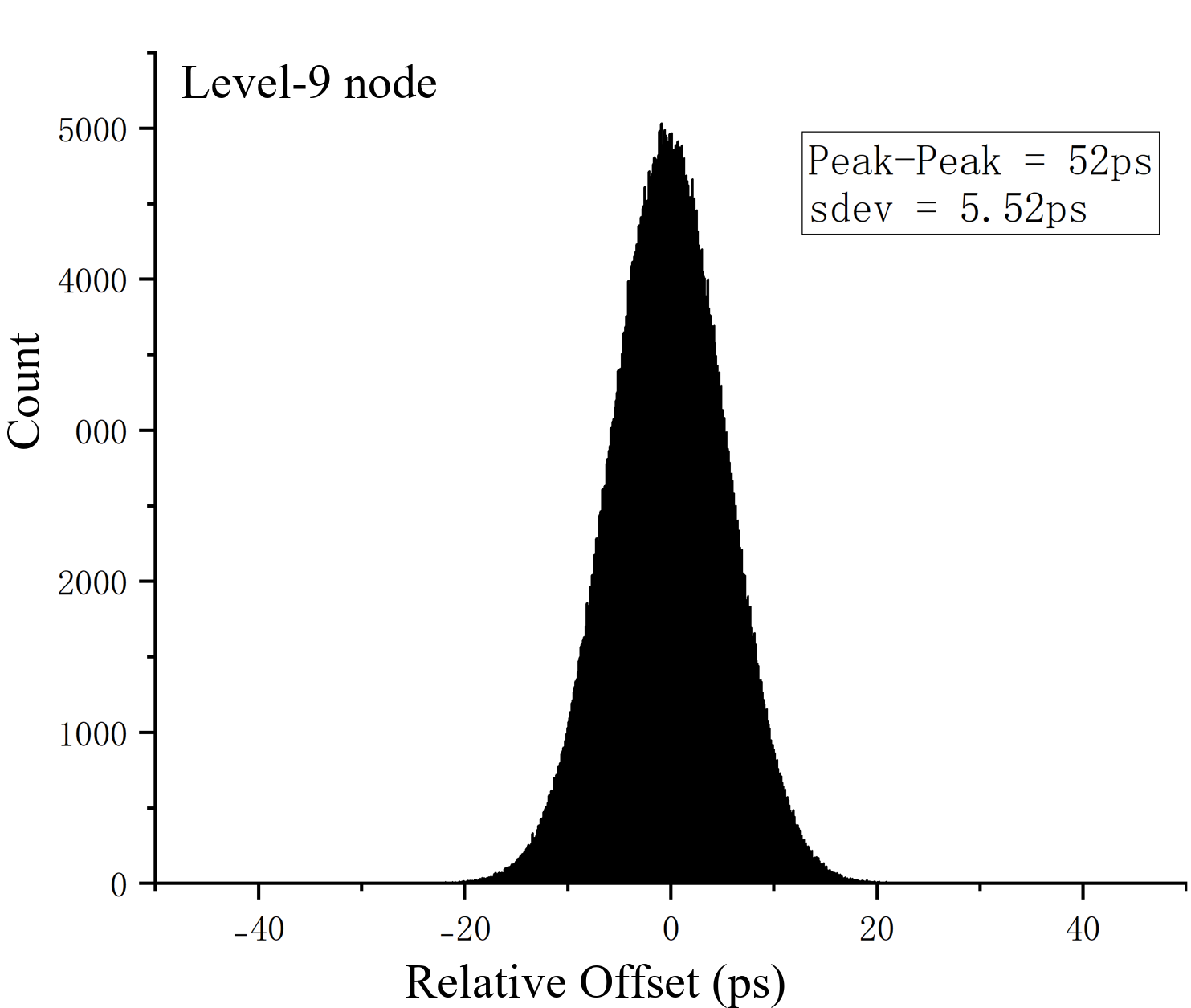}
        \caption{Level~9, $\sigma=5.52$~ps}
    \end{subfigure}
    \begin{subfigure}[b]{0.32\textwidth}
        \includegraphics[width=\textwidth]{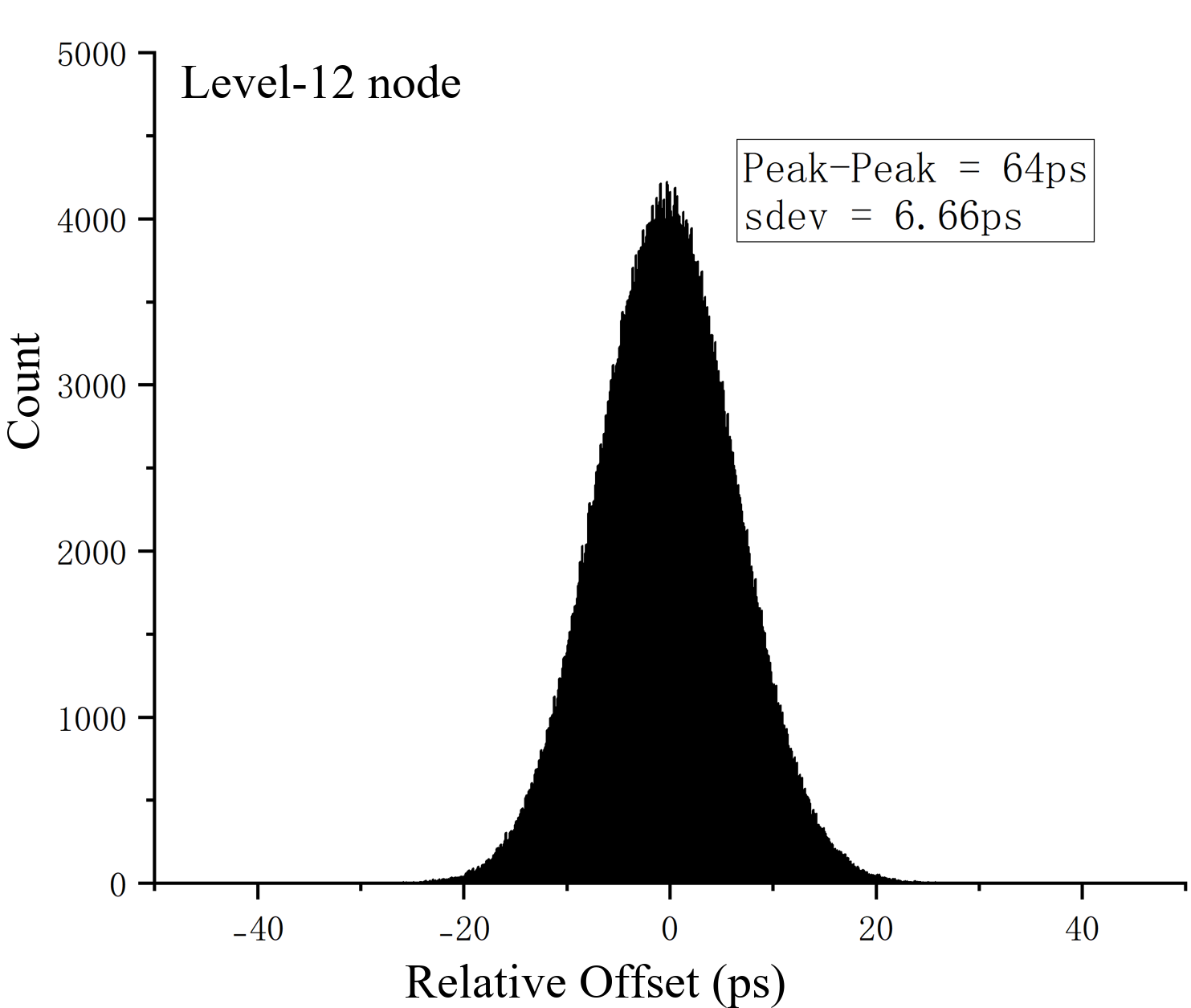}
        \caption{Level~12, $\sigma=6.66$~ps}
    \end{subfigure}
    \caption{12-level cascade results in a constant-temperature environment.}
    \label{fig:casc_const}
\end{figure}

\textbf{Constant-temperature results} (Fig.~\ref{fig:casc_const}): The synchronization precision at cascade levels 5, 9, and 12 is 4.63~ps, 5.52~ps, and 6.66~ps, respectively---all well within the 30~ps CEPC requirement.

\begin{figure}[htbp]
    \centering
    \includegraphics[width=0.5\textwidth]{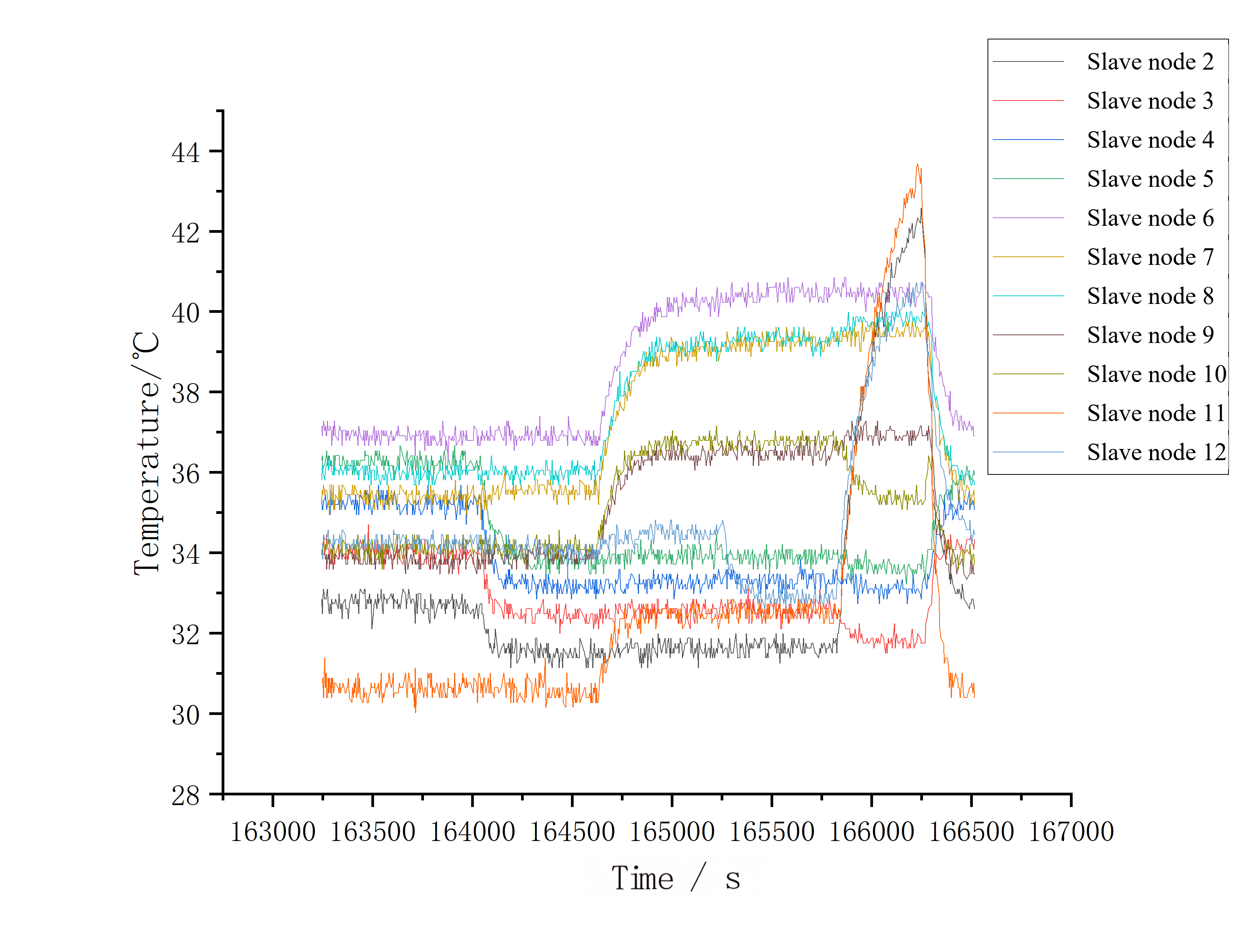}
    \caption{Per-node temperature trajectories during the variable-temperature test ($\sim$13$\,^\circ$C swing across nodes).}
    \label{fig:temp_curve}
\end{figure}

\begin{figure}[htbp]
    \centering
    \begin{subfigure}[b]{0.32\textwidth}
        \includegraphics[width=\textwidth]{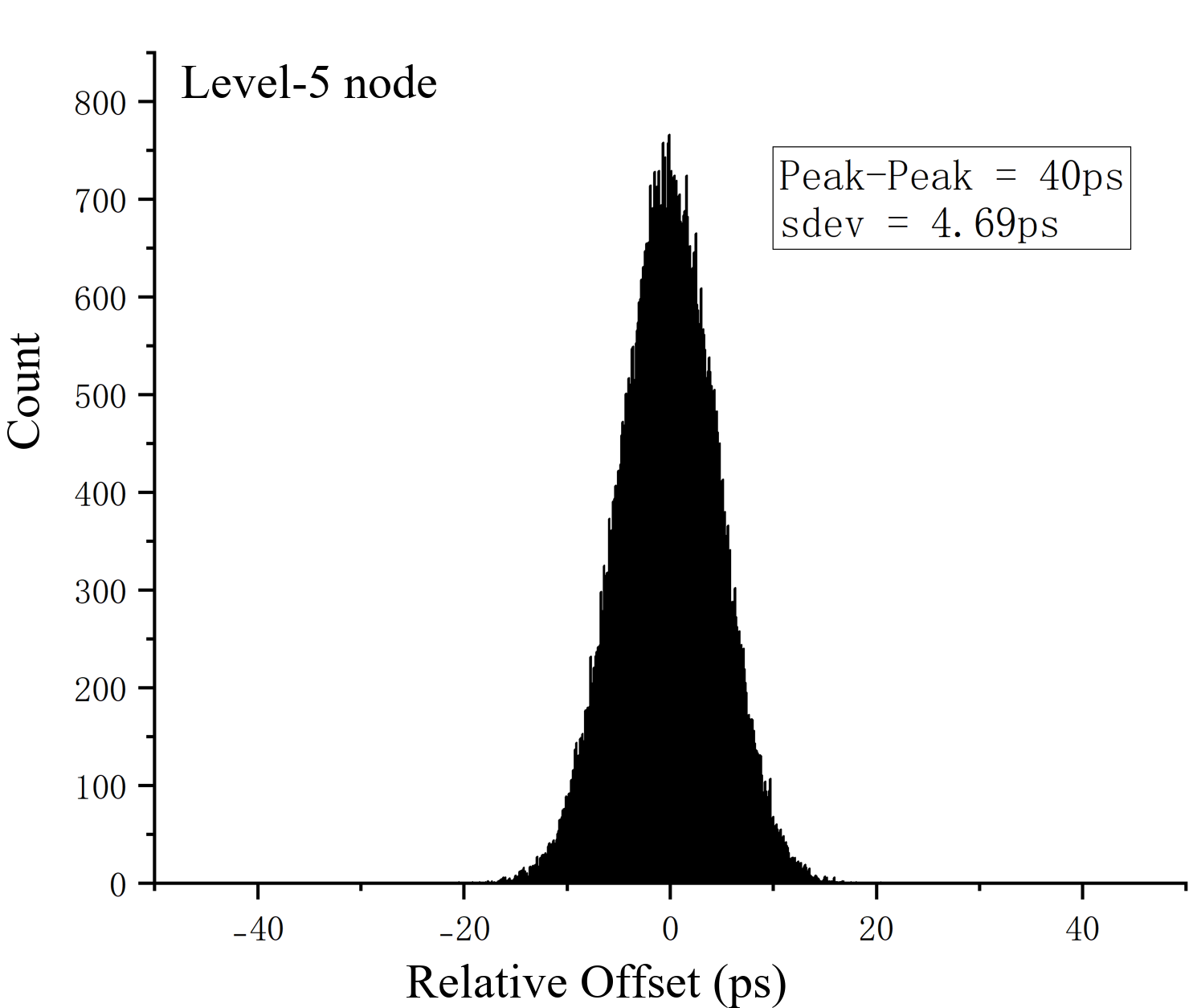}
        \caption{Level~5, $\sigma=4.69$~ps}
    \end{subfigure}
    \begin{subfigure}[b]{0.32\textwidth}
        \includegraphics[width=\textwidth]{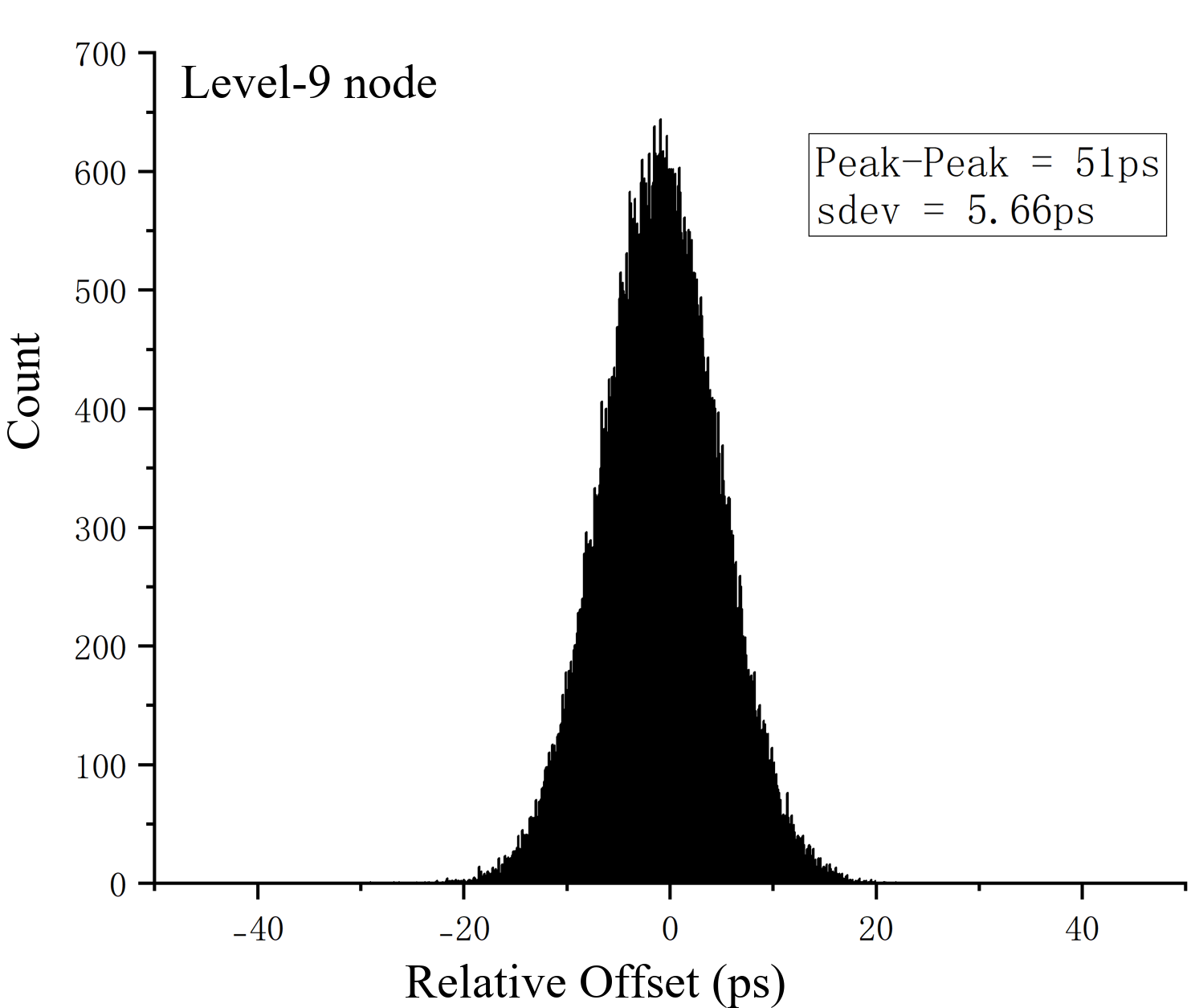}
        \caption{Level~9, $\sigma=5.66$~ps}
    \end{subfigure}
    \begin{subfigure}[b]{0.32\textwidth}
        \includegraphics[width=\textwidth]{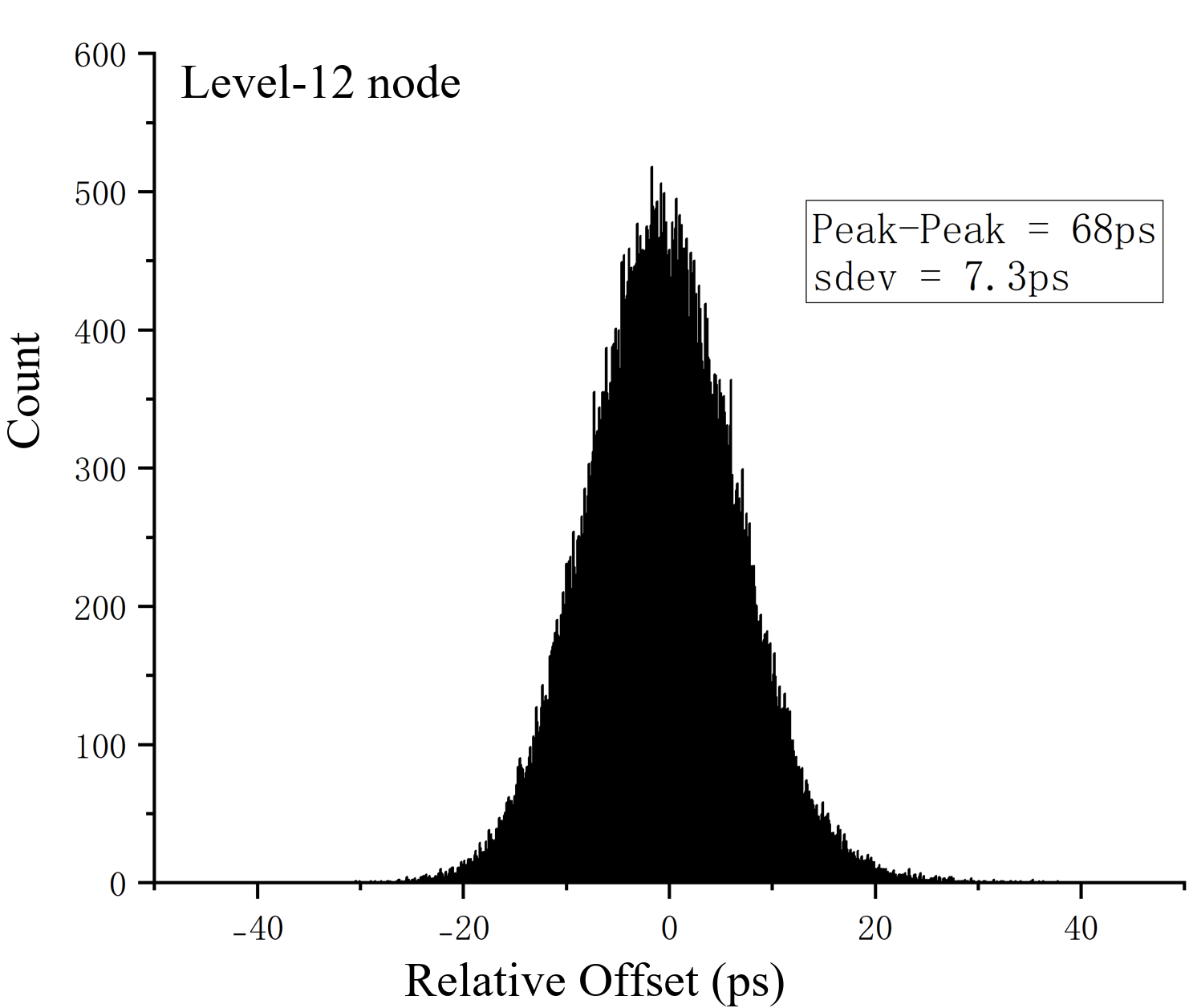}
        \caption{Level~12, $\sigma=7.30$~ps}
    \end{subfigure}
    \caption{12-level cascade results under the variable-temperature profile of Fig.~\ref{fig:temp_curve}.}
    \label{fig:casc_var}
\end{figure}

\textbf{Variable-temperature results}: Environmental temperature variation was emulated by switching on/off a set of fans pointed at the nodes, producing the per-node temperature trajectories shown in Fig.~\ref{fig:temp_curve} with a $\sim$13$\,^\circ$C amplitude over $\sim$1~h. Under these conditions the precision at levels~5, 9, and~12 became 4.69~ps, 5.66~ps, and 7.30~ps (Fig.~\ref{fig:casc_var}). At the deepest level (12), precision degrades by only 0.64~ps (from 6.66~ps to 7.30~ps) under a 13$\,^\circ$C inter-node temperature swing. This is consistent with the residual temperature coefficient of $-0.00425\,\mathrm{ps}/^\circ$C obtained in Section~\ref{subsec:tempcomp}, and confirms that the temperature feed-forward effectively cancels inter-node temperature drift.

\subsection{Long-Term Stability}

\begin{figure}[htbp]
    \centering
    \begin{subfigure}[b]{0.48\textwidth}
        \includegraphics[width=\textwidth]{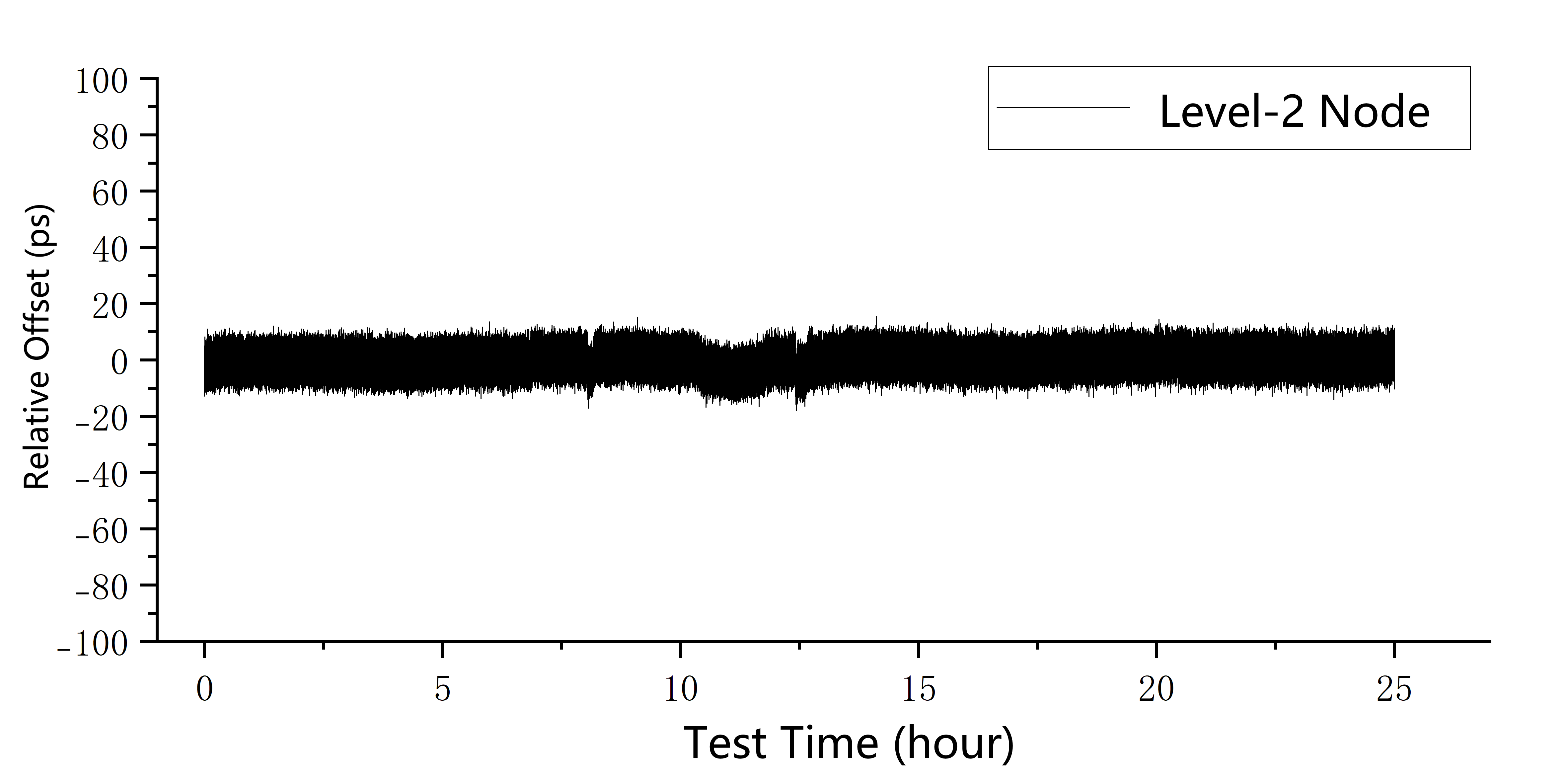}
        \caption{Node~2}
    \end{subfigure}
    \begin{subfigure}[b]{0.48\textwidth}
        \includegraphics[width=\textwidth]{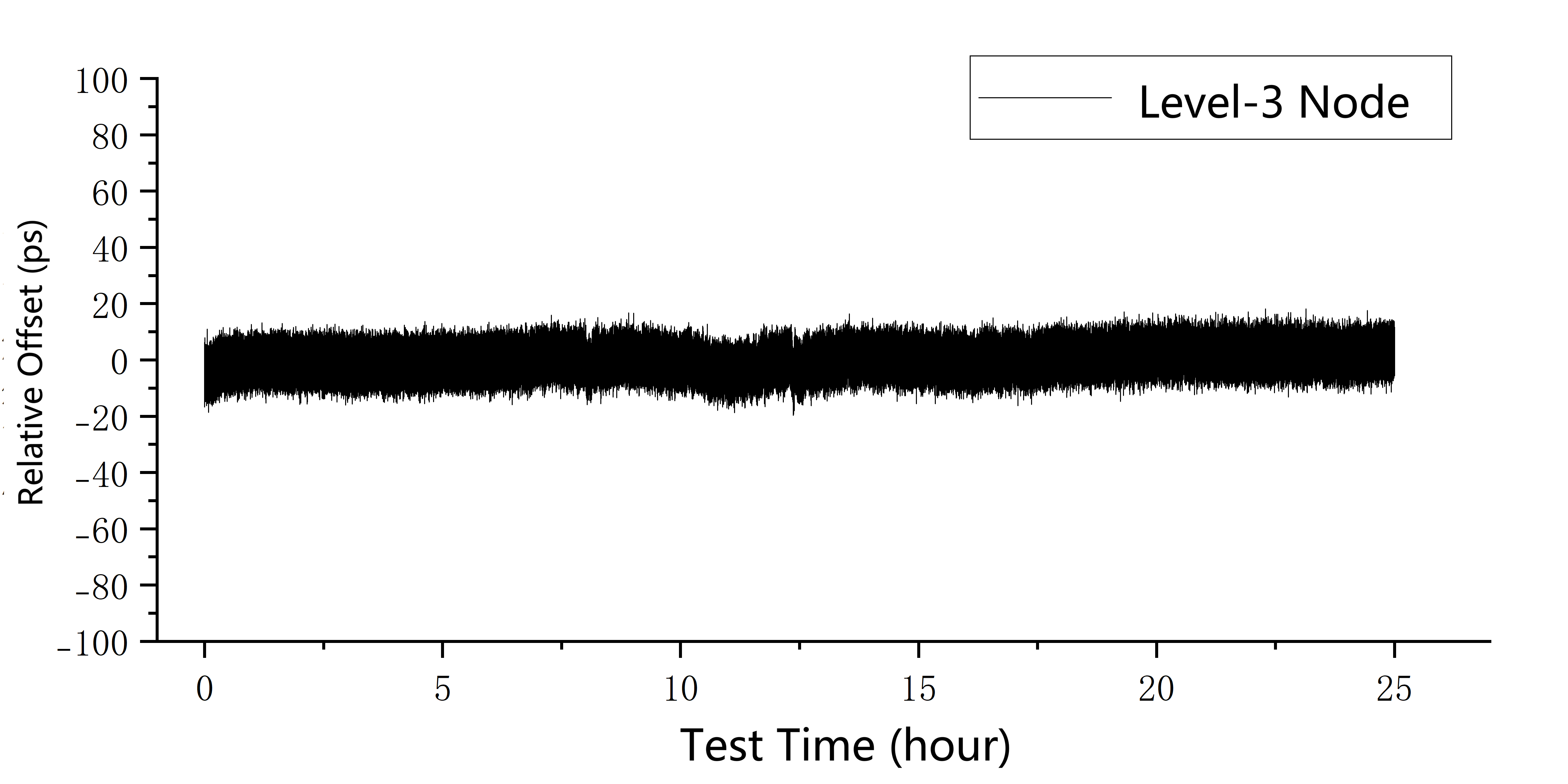}
        \caption{Node~3}
    \end{subfigure}
    \begin{subfigure}[b]{0.48\textwidth}
        \includegraphics[width=\textwidth]{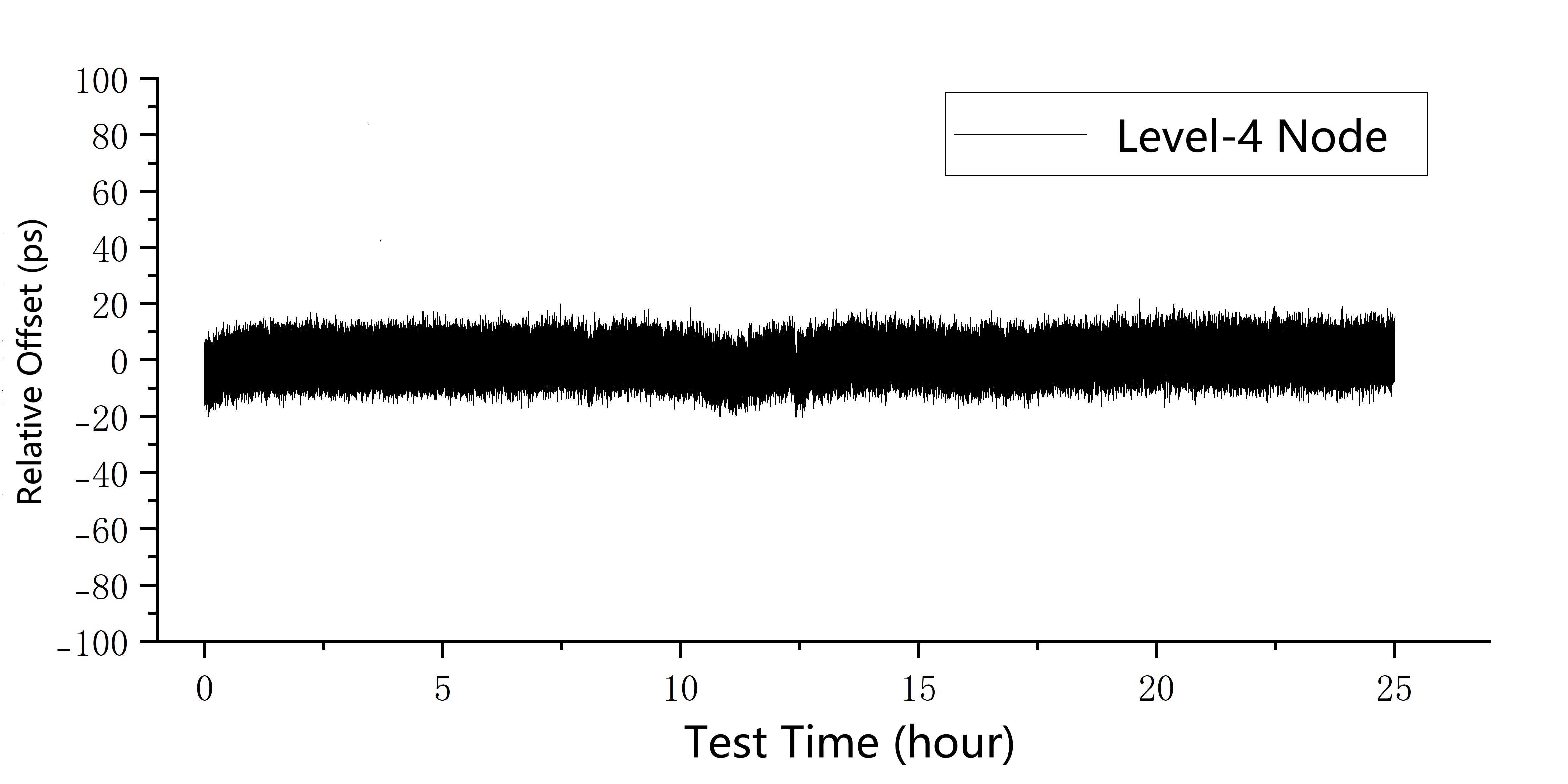}
        \caption{Node~4}
    \end{subfigure}
    \caption{25~h clock-offset trajectories for the 4-level cascade.}
    \label{fig:longterm_traj}
\end{figure}

A 4-level cascade was operated for 25~h at laboratory room temperature with global control active. Figure~\ref{fig:longterm_traj} shows the clock-offset trajectories of nodes 2, 3 and 4 relative to the master.

Over the full 25-h window no trend drift, lock loss, or sustained degradation was observed. The synchronization precision at nodes 2, 3 and 4 was 3.39~ps, 3.95~ps and 4.21~ps, respectively, demonstrating that level-to-level growth is modest even under continuous operation.

\subsection{Synchronized Clock Jitter Performance}

For the same 4-level cascade, the Time Interval Error (TIE) jitter, period jitter and cycle-to-cycle jitter were measured at all four nodes on scope~B using the on-instrument jitter package. All values are RMS over the acquisition window. Results are summarized in Table~\ref{tab:jitter}.

\begin{table}[htbp]
    \centering
    \caption{Synchronized-clock jitter (RMS) measured on scope~B on the 4-level cascade.}
    \label{tab:jitter}
    \begin{tabular}{cccc}
        \toprule
        Node & TIE (ps) & Period (ps) & Cycle-to-cycle (ps) \\
        \midrule
        Node 1 (master) & 0.915 & 1.562 & 2.696 \\
        Node 2 & 0.930 & 1.617 & 2.745 \\
        Node 3 & 0.771 & 1.528 & 2.645 \\
        Node 4 & 0.863 & 1.477 & 2.556 \\
        \bottomrule
    \end{tabular}
\end{table}

TIE jitter stays better than 1~ps across all four nodes and is essentially independent of cascade depth, confirming that the noise-suppression mechanisms of Section~\ref{sec:hw} act per-hop rather than cumulatively. The values are roughly an order of magnitude below the tens-of-picoseconds typical of standard WR.

\subsection{Restart Uncertainty Optimization Results}

\begin{figure}[htbp]
    \centering
    \begin{subfigure}[b]{0.4\textwidth}
        \includegraphics[width=\textwidth]{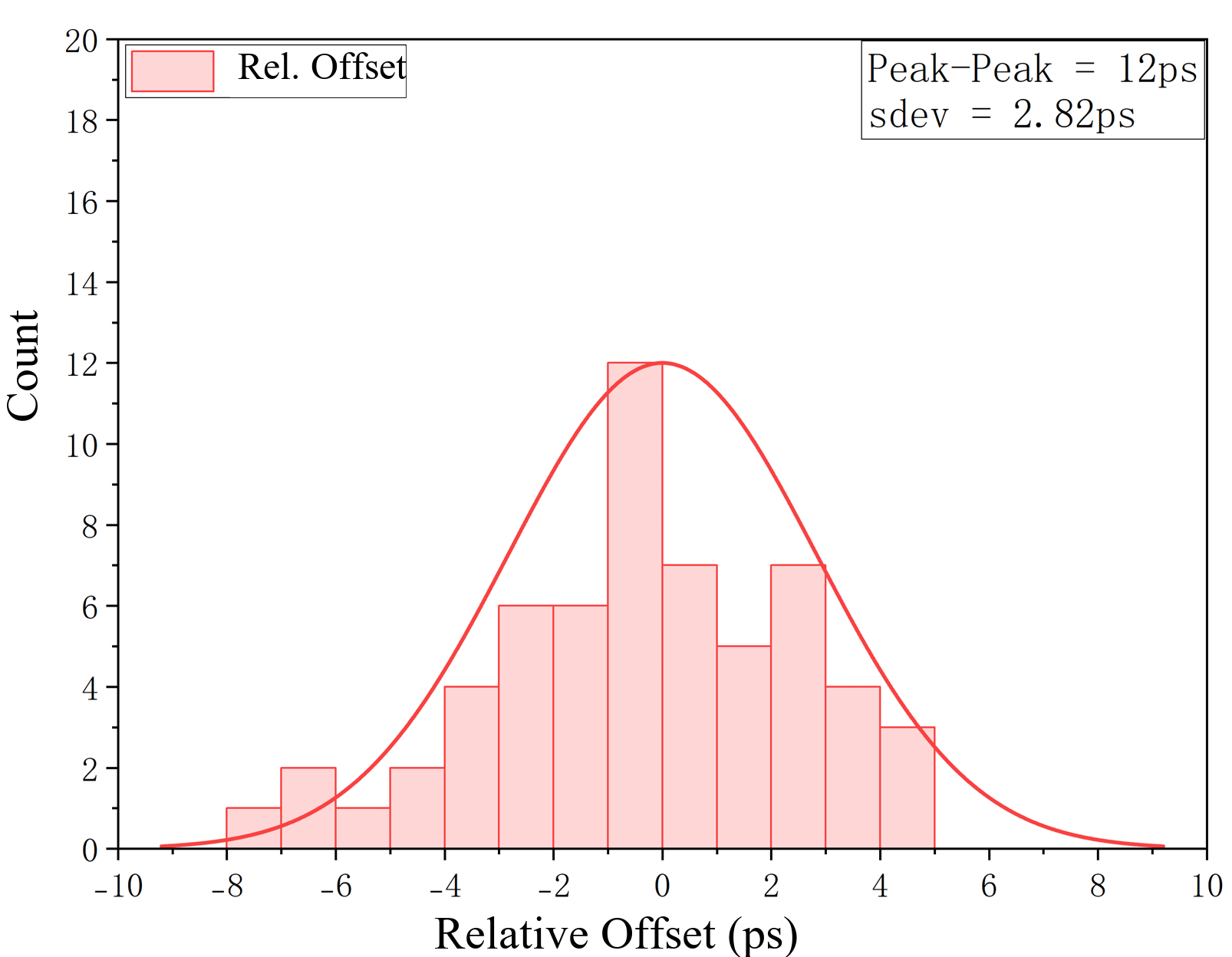}
        \caption{Histogram of mean offsets, $N=60$.}
    \end{subfigure}
    \begin{subfigure}[b]{0.4\textwidth}
        \includegraphics[width=\textwidth]{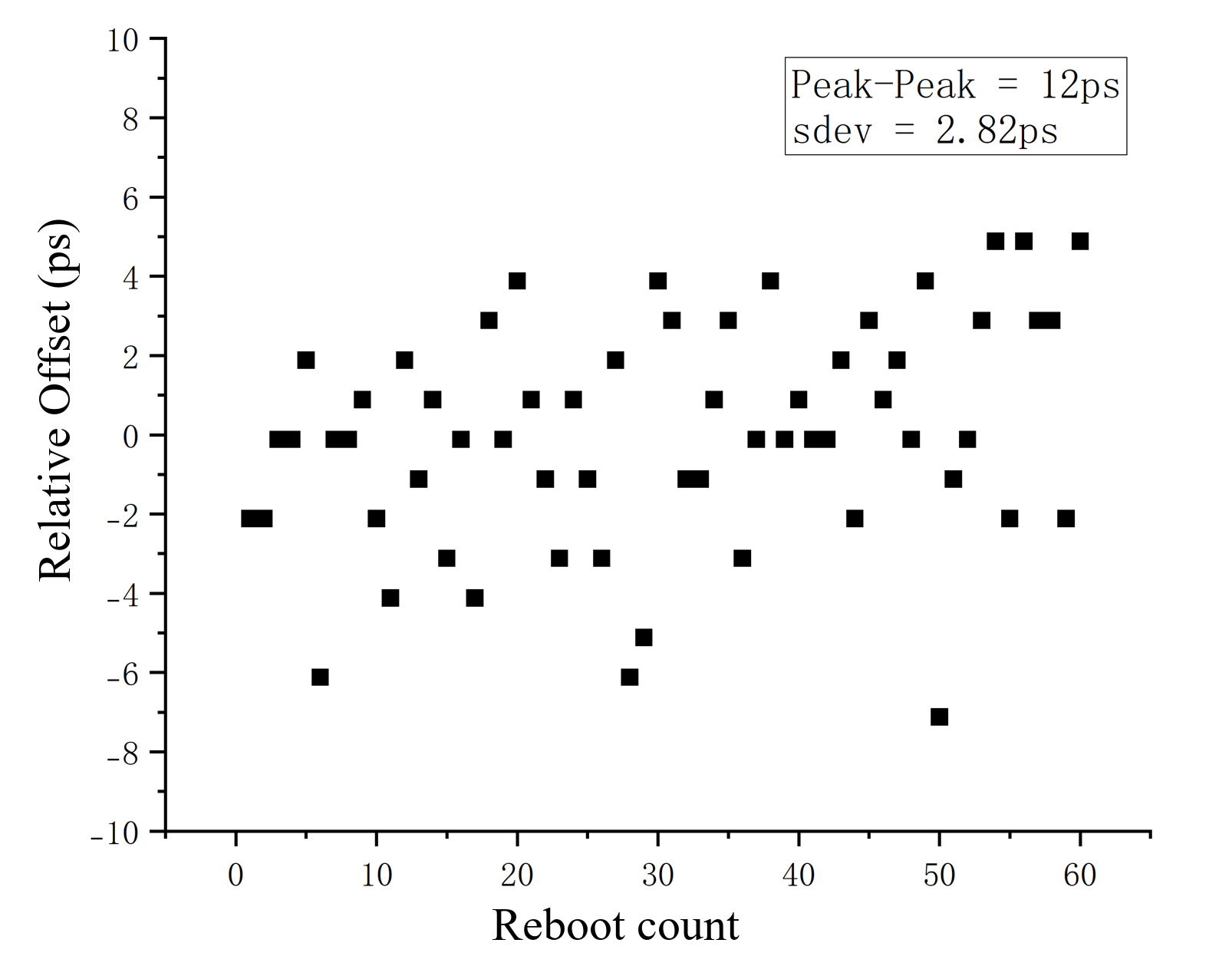}
        \caption{Restart-by-restart scatter plot.}
    \end{subfigure}
    \caption{Restart-uncertainty test: the master stays powered on while the slave is power-cycled 60 times. After each cycle, the system is given time to re-lock and settle (about 1--2~min); the steady-state mean offset is then recorded and averaged over $10^4$ samples on scope~B.}
    \label{fig:restart}
\end{figure}

Figure~\ref{fig:restart} reports 60 restart cycles with the optimized firmware. The peak-to-peak variation is 12~ps with a standard deviation of 2.82~ps, compared with the 88.8~ps peak-to-peak and 14.7~ps standard deviation of the standard WR baseline~\cite{ref_new87}. This is a 7.4$\times$ reduction in peak-to-peak and a 5.2$\times$ reduction in standard deviation, confirming the effectiveness of the GTX phase-alignment and byte-alignment-fixing optimizations described in Section~\ref{sec:fw}.

\subsection{Summary of Results and Comparison with the Standard-WR Baseline}

Table~\ref{tab:summary} summarizes the key performance metrics of the enhanced system and compares them with the standard-WR baseline reported in~\cite{ref_new87}.

\begin{table}[htbp]
    \centering
    \caption{Summary of the enhanced system and comparison with the standard-WR baseline reported in~\cite{ref_new87} and with the CEPC requirement.}
    \label{tab:summary}
    \begin{tabular}{lccc}
        \toprule
        Metric & This work & Standard WR (ref.) & CEPC requirement \\
        \midrule
        Point-to-point precision (1~m fiber)        & 3.38~ps & $\sim$20--30~ps & 30~ps \\
        Point-to-point precision (50~km fiber)      & 3.92~ps & ---             & 30~ps \\
        12-level cascade precision (const.\ temp.)  & 6.66~ps & ---             & 30~ps \\
        12-level cascade precision (13$\,^\circ$C var.) & 7.30~ps & ---         & 30~ps \\
        4-level 25~h precision (worst node)         & 4.21~ps & ---             & 30~ps \\
        TIE jitter (RMS)                            & <1~ps & tens of ps        & --- \\
        Restart uncertainty (std.\ dev.)            & 2.82~ps & 14.7~ps         & --- \\
        Restart uncertainty (peak-to-peak)          & 12~ps  & 88.8~ps          & --- \\
        \bottomrule
    \end{tabular}
\end{table}

All measured precision metrics meet the 30~ps CEPC requirement with at least a $\sim$4$\times$ margin in the worst case (12-level cascade under temperature variation). Where direct comparison is possible, the enhanced system outperforms the standard WR baseline by factors of 5 to 10.

\section{Conclusion}\label{sec:conclusion}
% Section 7: Conclusion

We have described an enhanced WR-based clock synchronization system that meets the 30~ps requirement of the CEPC accelerator. Beyond the immediate application, this work offers three contributions relevant to large-scale accelerator timing:

First, replacing the standard DAC+VCXO actuation chain with a Si5345A DSPLL+DCO restructures the WR slave loop into an inner fast-locking PLL realized entirely inside the clock generator and an outer slow (1~s) phase-setpoint correction driven by WR-PTP. This removes the board-level analog tuning node that dominates the close-in actuation noise and can be applied to any WR endpoint that needs sub-tens-of-ps synchronization.

Second, restart uncertainty---historically a $\sim$90~ps peak-to-peak bottleneck for WR---is reduced to 12~ps peak-to-peak (2.82~ps std.\ dev.) by combining GTX TX/RX phase alignment, manual byte-alignment fixing, and TXOUTCLK-vs-local-clock phase fixing during initialization. None of these steps require additional hardware; they only require an initialization sequence that takes less than 1~min on average and is acceptable for accelerator operation.

Third, the global-control architecture demonstrates that a PC-side controller with RL-trained gains and temperature feed-forward can keep the level-to-level precision growth essentially flat across a 12-level cascade ($\sim$6.66~ps), even under 13$\,^\circ$C of inter-node temperature variation. By keeping the online controller a fixed linear PID and using RL only to obtain its gains, we retain the determinism and analyzability that accelerator operators require, while letting the training stage absorb the complexity of multi-node, multi-parameter tuning.

Future work points in three directions. (i)~A field campaign under conditions closer to the actual CEPC tunnel---kilometer-scale per-hop fiber spans, distributed temperature gradients and mechanical vibration---is needed to confirm that the laboratory results extrapolate to the deployed topology. (ii)~The global-control framework will be extended toward distributed DDS-based multi-frequency RF clock distribution and toward online multi-agent learning, so that the controller can keep adapting during long accelerator runs without operator intervention. (iii)~For the most stringent determinism requirements, the controller can be migrated from a general-purpose PC to an embedded FPGA or an FPGA+AI-accelerator platform inside the master node, removing OS-level latency without changing the data path validated in this work.

% --- Acknowledgments ---
\acknowledgments
This work was supported by the National Key Programme for S\&T Research and Development (Grant No.:
2023YFA1606300).

% --- Bibliography ---
% For final JINST submission, replace `unsrtnat` with `JHEP`:
%   \bibliographystyle{JHEP}
% This requires JHEP.bst to be available (included in the JINST template package).
\bibliographystyle{JHEP}
\bibliography{ref}

\end{document}